\documentclass[a4paper,fleqn]{cas-dc}

\usepackage[authoryear,longnamesfirst]{natbib}
\usepackage{multirow}
\usepackage{graphicx}

\usepackage[linesnumbered,ruled,vlined]{algorithm2e}

\usepackage{caption}
\usepackage{hyperref}

\usepackage[most]{tcolorbox} 
\definecolor{ash}{rgb}{0.7, 0.7, 0.7}  
\definecolor{lightash}{rgb}{0.9, 0.9, 0.9}  

\usepackage{xcolor}
\usepackage{xurl} 
\usepackage[most]{tcolorbox} 
\usepackage{multicol}

\usepackage{pdflscape}
\usepackage{adjustbox} 
\usepackage[utf8]{inputenc}
\usepackage{float}
\usepackage{placeins}

\def\tsc#1{\csdef{#1}{\textsc{\lowercase{#1}}\xspace}}
\tsc{WGM}
\tsc{QE}
\tsc{EP}
\tsc{PMS}
\tsc{BEC}
\tsc{DE}

\begin{document}
\let\WriteBookmarks\relax
\def\floatpagepagefraction{1}
\def\textpagefraction{.001}

\shorttitle{MLRan: A Behavioural Dataset for Ransomware Analysis and Detection}

\shortauthors{FC Onwuegbuche et~al.}

\title [mode = title]{MLRan: A Behavioural Dataset for Ransomware Analysis and Detection}                      



\author[1,2,3]{Faithful Chiagoziem Onwuegbuche}[type=editor,
                        orcid=0000-0001-9580-4260]

\cormark[1]


\ead{faithful.chiagoziemonwuegb@ucdconnect.ie}


\credit{Conceptualization, Data curation, Funding acquisition, Formal analysis, Investigation, Methodology, Resources, Software, Visualization, Validation, Writing – original draft, Writing – review \& editing, Project administration}

\affiliation[1]{organization={School of Computer Science, University College Dublin},
    city={Dublin},
    country={Ireland}}

\affiliation[2]{organization={SFI Center for Research Training in Machine Learning},
    city={Dublin},
    country={Ireland}}

    \affiliation[3]{organization={Lero, Research Ireland Centre for Software},
    country={Ireland}}

\author[4]{Adelodun Olaoluwa}[type=editor,
                        orcid= 0009-0009-2602-1479]
\credit{Software, Investigation}

\affiliation[4]{organization={University of Ibadan},
    country={Nigeria}}

\author[1]{Anca Delia Jurcut}[type=editor,
                        orcid=0000-0002-2705-1823]
\credit{Conceptualization, Methodology,  Supervision,  Writing - review \& editing}

\author[1,3]{Liliana Pasquale}[type=editor,
                        orcid= 0000-0001-9673-3054]
\credit{Conceptualization, Methodology,  Supervision, Writing - review \& editing}

\cortext[cor1]{Corresponding author}



\begin{abstract}
Ransomware remains a critical threat to cybersecurity, yet publicly available datasets for training machine learning-based ransomware detection models are scarce and often have limited sample size, diversity, and reproducibility. In this paper, we introduce MLRan, a behavioural ransomware dataset, comprising over 4,800 samples across 64 ransomware families and a balanced set of goodware samples. The samples span from 2006 to 2024 and encompass the four major types of ransomware: locker, crypto, ransomware-as-a-service, and modern variants. We also propose guidelines (GUIDE-MLRan), inspired by previous work, for constructing high-quality behavioural ransomware datasets, which informed the curation of our dataset.  
We evaluated the ransomware detection performance of several machine learning (ML) models using MLRan. For this purpose, we performed feature selection by conducting mutual information filtering to reduce the initial 6.4 million features to 24,162, followed by recursive feature elimination, yielding 483 highly informative features. The ML models achieved an accuracy, precision and recall of up to 98.7\%, 98.9\%, 98.5\%, respectively. Using SHAP and LIME, we identified critical indicators of malicious behaviour, including registry tampering,  strings, and API misuse. The dataset and source code for feature extraction, selection, ML training, and evaluation are available publicly to support replicability and encourage future research, which can be found at \url{https://github.com/faithfulco/mlran}.

\end{abstract}

 \begin{graphicalabstract}
    \centering
    \includegraphics[width=1\textwidth]{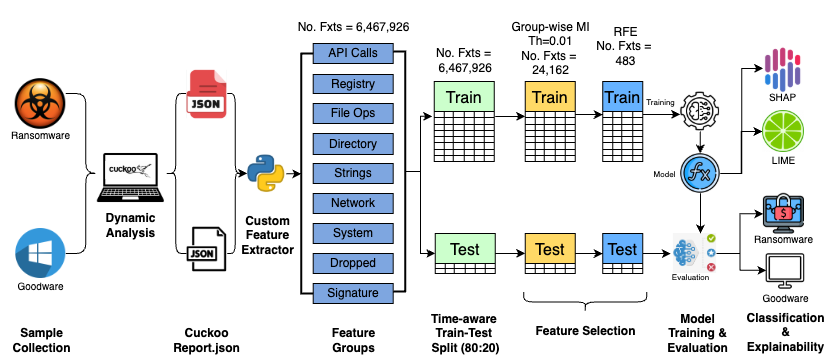}
 \end{graphicalabstract}

\begin{highlights}

\item MLRan: largest open-source behavioural ransomware dataset (64 families, 4.8K+ samples).

\item GUIDE-MLRan provides standardised guidelines for reproducible dataset creation.

\item Two-stage feature selection reduced 6.4M features to 483 without accuracy loss.


\item SHAP and LIME reveal key ransomware behaviours: strings, registry, and API.

\item Fully open-source pipeline: sandboxing, code for feature extraction and selection, ML training, and dataset.
\end{highlights}

\begin{keywords}
ransomware dataset \sep dynamic analysis \sep machine learning \sep malware \sep cybersecurity
\end{keywords}

\maketitle

\section{Introduction}
\label{sec:intro}

Ransomware attacks continue to pose a significant threat to individuals, organisations, and governments worldwide, causing substantial financial and operational disruptions \citep{onwuegbuche2023enhancing, beaman2021ransomware, jeremiah2024leveraging}. Recent statistics showcase the severity of the threat: in 2023, nearly seven out of every ten reported cyberattacks were ransomware-based, with over 317 million attempts recorded \footnote{Statista: \url{https://www.statista.com/topics/4136/ransomware/topicOverview}}. The financial impact is equally staggering, with the average cost per incident reaching \$1.85 million \footnote{Astra: \url{https://www.getastra.com/blog/security-audit/ransomware-attack-statistics/}}, and projections suggest that by 2031, global damage costs could soar to \$265 billion annually, with an attack occurring every two seconds \footnote{Cybercrime Magazine: \url{https://cybersecurityventures.com/ransomware-will-strike-every-2-seconds-by-2031/}}. These alarming figures not only illustrate why ransomware is now regarded as the foremost cyber threat by 62\% of C-suite executives \footnote{CFO: \url{https://www.cfo.com/news/cybersecurity-attacks-generative-ai-security-ransom/692176/}} but also highlight the urgent need for advanced detection and mitigation strategies.

Defending against ransomware remains challenging, as traditional signature-based antivirus measures are often insufficient \citep{mcintosh2021ransomware, kapoor2021ransomware}. In response, the cybersecurity community has increasingly turned to machine learning (ML) as a primary tool for ransomware detection and prevention \citep{al2024securing, onwuegbuche2023enhancing}. ML models can automatically learn complex behavioural patterns from large datasets, enabling them to recognise subtle indicators of ransomware activity—such as mass file encryption or abnormal process behaviour—that may elude human analysts or static rule-based systems \citep{fernando2024fesad}. Unlike signature-based approaches, ML-based methods can adapt to new ransomware types by (re-)training on updated data, keeping pace with the constantly evolving tactics of attackers.

However, the effectiveness of these ML models depends on the quality and representativeness of the data on which they are trained \citep{herrera2023dynamic}. Well-trained algorithms are only as robust as the datasets that underpin their learning. Most research surveys on ML-based ransomware detection have identified the limited availability of publicly accessible, high-quality datasets as a significant challenge \citep{ispahany2024ransomware, beaman2021ransomware, mcintosh2021ransomware, alraizza2023ransomware}. As noted by \cite{urooj2021ransomware}, no benchmark dataset currently serves as a reliable foundation for developing and evaluating ransomware detection systems. 
Several earlier datasets \citep{sgandurra2016automated, herrera2023dynamic, jethva2020multilayer, hirano2022ransap, hirano2025ransmap} feature small sample sizes and cover only a limited range of ransomware families, restricting their representativeness and generalisation capabilities. In addition, many datasets are imbalanced, often containing disproportionately fewer benign samples compared to ransomware or vice versa, thus skewing ML model training \citep{hou2024empirical, continella2016shieldfs, marcinkowski2024mirad}. Furthermore, most datasets capture only a narrow aspect of ransomware behaviour—such as low-level I/O operations, storage and memory access patterns \citep{continella2016shieldfs, hirano2022ransap, hirano2025ransmap}—without capturing comprehensive behavioural features. Manual and inconsistent data curation processes further impede reproducibility and the standardisation of experimental conditions, while the lack of public availability hinders future research \citep{moreira2024comprehensive}.


Our approach addresses these limitations by proposing MLRan, a large, diverse, balanced behavioural ransomware dataset featuring over 4800 balanced samples of ransomware and goodware. Our dataset focuses on ransomware targeting Windows systems since 95\% of the considered ransomware files are Windows-based executables or DLLs \footnote{VirusTotal Report on Ransomware in the Global Context: \url{https://blog.virustotal.com/2021/10/ransomware-in-global-context.html}}. MLRan covers 4 major ransomware types, namely locker, crypto,
ransomware-as-a-service (RaaS), and modern variants. It includes 64 ransomware families from 2008 to 2024, and captures nine key behavioural features, including API calls, registry keys, file and directory operations, strings, network activity, system processes, dropped files, and digital signatures. 

Inspired by previous work, we propose guidelines (GUIDE-MLRan) for constructing high-quality behavioural ransomware datasets and ensuring their reproducibility. We used these guidelines to ensure the rigorous curation of the MLRan dataset.  To streamline the dynamic analysis of malware necessary to curate the dataset, we enhanced the functionality of the Cuckoo Sandbox (an open-source automated malware analysis system) by automating file submission and sorting of analysis results. This extension reduced the manual effort necessary to generate the dataset, ensuring its consistent and efficient development. 

We empirically demonstrate that the MLRan dataset is well-suited for building effective ransomware detection models. To reduce data dimensionality,  we implemented a novel feature selection strategy based on mutual information filtering to reduce the initial 6.4 million features to 24,162, followed by recursive feature elimination, yielding 483 highly informative features.  The ML models achieved an accuracy, precision and recall of up to 98.7\%, 98.9\%, 98.5\%, respectively. Our feature selection strategy reduces data dimensionality and computational cost without compromising performance, highlighting the dataset’s practicality and scalability.
 
 We analyse key ransomware behaviours using explainable AI techniques such as SHapley Additive exPlanations (SHAP) and Local Interpretable Model-agnostic Explanations (LIME). This analysis identifies the most discriminative features for distinguishing ransomware from benign software, providing valuable insights for future research on ransomware detection.
We made our dataset and tool implementation publicly available \footnote{MLRan GitHub Repository: \url{https://github.com/faithfulco/mlran} } to enhance reproducibility and support future research. The tool provides several functionalities, including ransomware and goodware sample collection, automated file processing and results sorting in Cuckoo Sandbox, feature extraction for nine behavioural categories, feature selection using our proposed strategy, and machine learning code, enabling researchers to replicate and extend our work as new ransomware variants emerge. All data sharing follows strict ethical guidelines and excludes raw malware binaries.

The rest of the paper is organised as follows. In Section \ref{sec:backgroup}, we review existing publicly available behavioural ransomware datasets, their strengths and limitations and motivate the need for the MLRan dataset. In Section \ref{sec:guide_mlran} we introduce GUIDE-MLRan, our set of guidelines for developing high-quality behavioural ransomware datasets, while in Section \ref{sec:mlran_dataset}, we detail the proposed MLRan dataset. In Section \ref{sec:experimental_design}, we discuss the experimental design used to address the research questions, and in Sections \ref{sec:results}, we present the research questions' results and compare them with related works, respectively. Finally, in Section \ref{sec:conclusion}, we conclude the paper and provide recommendations for future research.


\section{Ransomware behavioural datasets and their limitations} \label{sec:backgroup}
This section presents a curated list of publicly available ransomware datasets suitable for training machine learning models and discusses their characteristics, contributions and limitations.

\subsection{Ransomware datasets}

 The \textbf{EldeRAN} ransomware dataset\footnote{Elderan: \url{https://rissgroup.org/ransomware-dataset/}} was published in 2016~\citep{sgandurra2016automated}. It comprises 582 ransomware samples spanning 11 families, along with 942 samples of goodware applications. Using the Cuckoo Sandbox on Windows XP 32-bit, EldeRAN dynamically analyses both ransomware and goodware to extract a rich set of 30,967 features—including API calls, registry operations, file system and directory activities, file extensions, dropped files, and embedded strings. A feature selection algorithm then identifies the most relevant features used in a Regularized Logistic Regression classifier to distinguish between ransomware and goodware. The ransomware samples were randomly selected from VirusShare \footnote{VirusShare: \url{http://virusshare.com/}}, and the goodware samples were obtained from a software aggregator website \footnote{Software Informer: \url{https://software.informer.com/software/}}, ensuring a diverse set of benign applications.

Despite its widespread use, the Elderan dataset's sample size is relatively small and imbalanced, with 582 ransomware samples compared to 942 goodware samples. Moreover, the dataset predominantly comprises crypto-ransomware, lacking representation of newer ransomware types such as RaaS and modern multi-stage attacks. Consequently, it may not adequately reflect the current threat landscape, limiting its utility for developing robust, up-to-date detection models.


The \textbf{ShieldFS} dataset\footnote{ShieldFS: \url{http://shieldfs.necst.it/}} is a self-healing, ransomware-aware filesystem that intercepts low-level I/O Request Packets (IRPs) to detect and mitigate ransomware attacks in real time~\citep{continella2016shieldfs}. The system was established by first developing IRPLogger, a custom kernel-level sniffer installed on volunteer machines to collect a comprehensive baseline of normal filesystem activity over one month—resulting in about 1.7 billion IRPs from 2,245 different applications. Subsequently, known ransomware samples—383 active instances from 5 families (CryptoWall, TeslaCrypt, Critroni, CryptoDefense, Crowti)—were executed on controlled machines configured with realistic decoy files and user data. By monitoring metrics such as file reads, writes, directory listings, and entropy changes, ShieldFS detects suspicious encryption-like behaviours and automatically rolls back malicious modifications. Although effective for mitigating file-encryption attacks, the dataset primarily captures file I/O patterns, lacks behavioural feature diversity (e.g., API calls and registry activity), and does not cover newer ransomware types such as RaaS and modern ransomware.

The \textbf{ISOT ransomware detection dataset}\footnote{ISOT: \url{https://onlineacademiccommunity.uvic.ca/isot/2022/11/27/botnet-and-ransomware-detection-datasets/}} was published in 2020 \citep{jethva2020multilayer}. This dataset was designed to capture a diverse range of ransomware families and variants, comprising 666 ransomware samples alongside 103 goodware samples. Data was collected using the Cuckoo Sandbox in a controlled Windows 7 environment to simulate real-world conditions as closely as possible. The dataset includes 51,556 features grouped into eight categories: API calls, registry key operations, command line operations, DLLs, directory enumerations, mutexes, embedded strings, and miscellaneous binary attributes. Despite its strengths in feature diversity, the relatively small number of samples (769 samples) and imbalanced distribution of goodware samples, when compared to ransomware, may skew model training.

The \textbf{RanSAP} dataset\footnote{RanSAP: \url{https://github.com/manabu-hirano/RanSAP/}} is an open dataset of ransomware storage access patterns designed for training machine learning models~\citep{hirano2022ransap}. It captures time-series data from 7 prominent ransomware samples alongside 5 benign software samples. Using a thin hypervisor-based monitoring system, RanSAP records low-level storage interactions, such as sector-based read and write operations and entropy measurements, across various operating conditions. The dataset encompasses samples executed under different configurations, including operating system versions and storage device configurations (e.g., devices with full drive encryption enabled). By focusing on storage access patterns, RanSAP provides insight into the dynamic behaviours of ransomware that are critical for developing detection models. However, the dataset scope is confined to storage-related features, omitting other behavioural features such as API calls and registry activities that may also be valuable for comprehensive detection. Additionally, the relatively small sample size restricts its representativeness of the broader ransomware ecosystem, as the ransomware families considered are just crypto and RaaS, omitting locker and newer modern ransomware types.


The \textbf{Dynamic Feature Dataset}\footnote{Dynamic Feature Dataset: \url{https://github.com/Juan-Herrera-Silva/Paper-SENSORS/tree/main}} is built from experiments on 40 artifacts (20 ransomware and 20 goodware) executed ten times on each of five victim devices, resulting in a total of 2000 experiments using the Cuckoo sandbox \citep{herrera2023dynamic}. The dataset comprises 50 distinct features organised into 7 feature groups, which capture dynamic behaviours such as API call flows, registry operations, file and directory activities, network communications, and other relevant runtime characteristics. Despite its balanced sample distribution and good experimental design, which leads to generating more observations, the dataset sample size is relatively modest compared to more extensive collections, which might restrict its representativeness across the full spectrum of ransomware types.

The \textbf{MarauderMap} ransomware dataset\footnote{MarauderMap: \url{https://github.com/m1-llie/MarauderMap-code}} comprises 7,796 active Windows ransomware samples spanning 95 families~\citep{hou2024empirical}. The samples were dynamically analysed using a custom testbed developed by the authors. During execution, the testbed collected 1.98TiB of runtime logs spanning six key behavioural categories: API calls, I/O operations, network traffic, registry modifications, command executions, and service management.
Samples were sourced from platforms like VirusTotal and VX Vault, with the majority identified between 2021 and 2023. Despite its comprehensive scale of ransomware samples and rich behavioural features, the dataset does not include goodware samples. While the study used benign executables for evaluating detection strategies, these were not part of the dataset and were not analysed in the sandbox. 

The \textbf{MIRAD} dataset\footnote{MIRAD: \url{https://github.com/Sagenso/MIRAD}}, introduced in 2024, is designed for dynamic ransomware detection~\citep{marcinkowski2024mirad}. It was generated by simulating ransomware attacks in a virtual environment that mimics typical office activity on Windows. The dataset includes 13 features extracted from event logs (e.g., API calls, registry changes, file events) aggregated over fixed intervals with a moving average. Each data point is labelled as “pre-attack” (benign) or “post-attack” (ransomware activity), comprising a total of 67,427 samples (39,440 for training and 27,987 for testing), and includes 78 ransomware samples. However, the small number of ransomware samples and features may not capture the full spectrum of real-world behaviours, and the documentation lacks critical metadata, which could hinder reproducibility and further research. 


The \textbf{Ransomware Combined Structural Feature (RCSF)} dataset\footnote{RCSF: \url{https://data.mendeley.com/datasets/yzhcvn7sj5}}  aggregates structural features from Windows PE files to support detection of emerging ransomware families through static analysis~\citep{moreira2024comprehensive}. It combines multiple features, including PE header fields, imported DLLs, function calls, section entropy, and opcodes of each binary sample. The dataset comprises 2675 samples, with a training set of 1023 ransomware (from 25 families) and 1134 goodware, and a testing set of 385 ransomware (from 15 families) and 133 goodware. Feature extraction is performed using tools such as the Python pefile library, Detect It Easy, and Distorm3, followed by variance threshold feature selection to refine the data. An ensemble model combining Logistic Regression, Random Forest, and XGBoost achieves over 97\% in accuracy, precision, recall, and F-measure, with a prediction time of about 0.37 seconds per sample. While the data includes rich structural features, it relies exclusively on static features extracted from Windows PE files. If ransomware employs obfuscation, packing, or encryption techniques, the structural features may be altered, potentially reducing detection accuracy.


\textbf{RanSMAP}\footnote{RanSMAP: \url{https://github.com/manabu-hirano/RanSMAP}} captures low-level storage and memory access patterns using a thin hypervisor~\citep{hirano2025ransmap}. Unlike its predecessor, RanSAP, which collected only storage access patterns, RanSMAP integrates memory access features to improve deep learning–based ransomware detection by about 2.3\%, especially under simultaneous execution of ransomware and benign applications. The RanSMAP dataset comprises 1,970 executions, which include 6 ransomware samples (WannaCry, Ryuk, REvil, LockBit, Darkside and Conti) and 6 goodware samples. Each execution is represented by a 23-dimensional feature vector, derived from two feature groups—5 storage access features and 18 memory access features. The dataset spans 6 distinct ransomware families and was collected on Windows 10. However, the dataset includes a relatively small number of ransomware samples and types, which may not fully represent real-world diversity across ransomware types.


Tables \ref{tab:datasets_description} and \ref{tab:datasets_characteristics} provide, respectively, a descriptive summary and key characteristics of major publicly available ransomware datasets, highlighting their methodologies, sample composition, and feature representation. Most datasets rely on dynamic analysis, with RCSF being the only exception that employs static analysis. Commonly used tools include Cuckoo Sandbox, BitVisor, and IRP Logger, with Cuckoo Sandbox being the most widely used. Some datasets, such as MarauderMap and MIRAD, utilise proprietary testbeds, which limit reproducibility.

The sources of ransomware samples vary, with VirusTotal being the most frequently used repository. Additional datasets leverage sources such as MalwareBazaar, Hybrid-Analysis, and VirusShare, while MarauderMap uniquely obtains some samples from dark web markets and hacker forums, offering potentially more diverse data. In terms of operating systems, most datasets focus on Windows 7 and Windows 10, with only a few using multiple Windows versions in analysing the samples, such as the Dynamic Feature dataset. MLRan, however, integrates Cuckoo Sandbox and obtains ransomware samples from multiple malware repositories, providing a more comprehensive and diverse dataset for analysis.

\subsection{Limitations}

One of the most significant limitations of existing datasets is their small size. Only MarauderMap contains 3,000 or more samples, while the rest remain relatively small. However, MarauderMap does not include goodware samples, making it unsuitable for training machine learning models that require both ransomware and benign software for accurate classification. Also, only 4 out of 9 datasets include a balanced representation of ransomware and goodware. MLRan addresses this limitation by offering a well-balanced dataset that comprises both ransomware and goodware, ensuring better training and generalisation for machine learning models.

Another key issue is the incomplete representation of ransomware types. No dataset comprehensively includes all ransomware types; datasets either prioritise recent ransomware at the expense of older samples or focus on past variants while neglecting modern threats. 8 out of 9 datasets include crypto ransomware, while locker and modern ransomware are significantly underrepresented. Additionally, 7 out of 9 datasets cover only up to 25 ransomware families, while MarauderMap includes 95 families, making it the most diverse. MLRan overcomes these gaps by providing full-spectrum coverage of historical and contemporary ransomware types across 64 families, ensuring better applicability in studying the evolution of ransomware attacks. 

Goodware representation is another critical limitation. Although 7 out of 9  datasets include goodware, only 3 datasets have a diverse selection of benign software. This imbalance reduces the effectiveness of machine learning models, as a well-balanced dataset is essential for distinguishing between malicious and benign software. MLRan effectively resolves this issue by incorporating a diverse and balanced set of goodware, enhancing its real-world applicability.

Feature representation across these datasets also varies significantly. File and directory operations are included in all the datasets, reflecting ransomware’s primary impact on file systems. However, API calls and registry keys are only included in 6 and 5 datasets, respectively, despite their importance in behavioural analysis. Additionally, network-based features, system processes, dropped operations, and signatures are rarely considered; that is, each feature is only included in two datasets, which limits the effectiveness of the datasets in detecting ransomware using dynamic behavioural indicators. MLRan incorporates all these critical features, ensuring a richer dataset that enhances machine learning-based ransomware detection.


MLRan addresses these shortcomings by offering a large, well-balanced dataset that includes diverse ransomware types, comprehensive goodware samples, and an extensive feature set, making it a superior dataset for advancing ransomware detection research. 

\begin{table*}[ht]
\caption{Descriptive summary of the major publicly available ransomware datasets. N denotes the total number of samples, \#Ran represents the number of ransomware samples, and \#Good indicates the number of goodware samples. \#Fx refers to the total number of features, while \#RanF specifies the number of ransomware families. \#FxG denotes the number of feature groups, and OS indicates the operating system used in the dataset. The symbol '-' indicates that the respective information was not found in the dataset/paper.}
\label{tab:datasets_description}
\resizebox{\textwidth}{!}{%
\begin{tabular}{|l|c|c|c|c|c|c|c|c|c|c|c|}  
\hline
\textbf{Dataset}                                                                           & \textbf{Year} & \textbf{Analysis} & \textbf{N} & \textbf{\#Ran} & \textbf{\#Good} & \textbf{\#Fx} & \textbf{\#RanF} & \textbf{\#FxG}& \textbf{\begin{tabular}[c]{@{}c@{}}Software\\ Used\end{tabular}}& \textbf{\begin{tabular}[c]{@{}c@{}} Sample\\ Source\end{tabular}}                                                                                       & \textbf{OS}                                                       \\ \hline
\textbf{Elderan }                                                                           & 2016          & Dynamic           &         1524        &      582             & 942                  & 30967               & 11                                                                        & 7                                                                     & \begin{tabular}[c]{@{}c@{}}Cuckoo\\ Sandbox\end{tabular}            & \begin{tabular}[c]{@{}c@{}}VirusShare\\ Software informer\end{tabular}                                                                                      & Win XP                                                                \\ \hline

\textbf{ShieldFS }                                                                           & 2016          & Dynamic           &    2628             &                  2245   & 383                  & 6               & 5                                                                        & 2                                                                     & \begin{tabular}[c]{@{}c@{}}IRPLogger,\\ ShieldFS\\ driver\end{tabular} ,          & VirusTotal                                                                                                    & Win 7                                                               \\ \hline

\textbf{\begin{tabular}[c]{@{}l@{}}ISOT \end{tabular}}              & 2020          & Dynamic           & 769                 & 666                    & 103                  & 51556                & 20                                                                        & 8                                                                     & \begin{tabular}[c]{@{}c@{}}Cuckoo\\ Sandbox\end{tabular}          & VirusTotal                                                                                                    & Win 7                                                             \\ \hline

\textbf{RanSAP }                                                                            & 2022          & Dynamic           & 12                  & 7                      & 5                    & 5                 & 7                                                                     & 1                                                                   & BitVisor               & ANY.RUN                                                                                                          & \begin{tabular}[c]{@{}c@{}}Win Server\\ 2008 R2 \\ and Win 7\end{tabular} \\ \hline

\textbf{\begin{tabular}[c]{@{}l@{}}Dynamic \\Feature \end{tabular}}                                                                  & 2023          & Dynamic           & 40                  & 20                     & 20                   & 50                   & 20                                                                      & 7                                                                     & \begin{tabular}[c]{@{}c@{}}Cuckoo\\ Sandbox\end{tabular}                & \begin{tabular}[c]{@{}c@{}}VirusShare, theZoo,\\ VirusTotal and\\ Hybrid-Analysis\end{tabular}              & \begin{tabular}[c]{@{}c@{}}5 Different\\ Windows OS\end{tabular}                                                         \\ \hline

\textbf{MarauderMap }                                                                       & 2024          & Dynamic           & 7,796                 & 7,796                  & -                  & -                  & 95                                                                        & 6                                                                     & \begin{tabular}[c]{@{}c@{}}Author\\ Testbed\end{tabular}            & \begin{tabular}[c]{@{}c@{}}VirusTotal, Vx vault,\\ hacker forums, \\ dark web, \\black market groups\end{tabular} & Win 10                                                                      \\ \hline

\textbf{MIRAD }                                                                       & 2024          & Dynamic           &        78         &    78               &          -        & 13                 & -                                                                        & 6                                                                     & \begin{tabular}[c]{@{}c@{}}Author\\ Testbed\end{tabular}         & MalwareBazaar & Win 10 \& 11                                                                     \\ \hline

\textbf{\begin{tabular}[c]{@{}l@{}}RCSF \end{tabular}} & 2024          & Static            & 2675                & 1408                   & 1267                 & 11027                & 25                                                                        & 4                                                                   & \begin{tabular}[c]{@{}c@{}}pefile lib\\ DIE \\ Distorm3\end{tabular}                    & \begin{tabular}[c]{@{}c@{}}VirusShare\\ Hybrid-Analysis\end{tabular}                                                                          & Windows                                                                         \\ \hline

\textbf{RanSMAP }                                                                            & 2025          & Dynamic           & 12                  & 6                      & 6                    & 23                  & 6                                                                       & 2                                                                  & BitVisor               & ANY.RUN                                                                                                           & Win 10 \\ \hline

\textbf{\begin{tabular}[c]{@{}l@{}}MLRan \\ Our Dataset \end{tabular}}                                                                        & 2025          & Dynamic           & 4880                & 2330                    & 2550                  & 6,467,926              & 64                                                                       & 9                                                                    & \begin{tabular}[c]{@{}c@{}}Cuckoo\\ Sandbox\end{tabular}          & \begin{tabular}[c]{@{}c@{}}VirusShare\\ VirusTotal\\ MalwareBazaar \\ Hybrid-Analysis \\ MarauderMap \end{tabular}                                                                                              & Win 7                                                              \\ \hline
\end{tabular}%
}
\end{table*}

\begin{table*}[!htp]
\centering
\caption{Key characteristics of major publicly available ransomware datasets. The presence of a specific characteristic is denoted by ‘$\checkmark$’ while its absence is marked as ‘$\times$’. The symbol '-' indicates that the respective information was not found in the dataset. A dataset is considered large if it contains 3,000 or more samples.  Within ransomware types, L represents locker ransomware, C denotes crypto ransomware, Rs signifies ransomware-as-a-service, and M indicates modern ransomware. For goodware, G denotes the presence of goodware samples, while D indicates if the goodware samples are diverse. Regarding labelling, B represents binary labelling (ransomware vs. goodware), F denotes family-based labelling, and T signifies type-based labelling. Feature groups represent the different feature groups available in the datasets. The different groups include API calls (API), registry keys (REG), file operations (FILE), directory operations (DIR), string-based features (STR), network-related features (NET), system processes (SYS), dropped operations (DROP), and signatures (SIG). Any additional features not falling into these categories are classified as Others. The row, '\% of $\checkmark$' indicates the percentage of datasets that have that characteristic, excluding our dataset, rounded to a whole number.}
\centering
\label{tab:datasets_characteristics}
\resizebox{\textwidth}{!}{%
\begin{tabular}{lccccc|cc|c|ccc|cccccccccc}
\hline
\multirow{2}{*}{\textbf{Dataset}} &
  \multirow{2}{*}{\textbf{\begin{tabular}[c]{@{}c@{}}Large\\ Size\end{tabular}}} &
  \multicolumn{4}{c|}{\textbf{Ranamware Types}} &
  \multicolumn{2}{c|}{\textbf{Goodware}} &
  \multicolumn{1}{c|}{\multirow{2}{*}{\textbf{Bal}}} &
  \multicolumn{3}{c|}{\textbf{Labelling}} &
  \multicolumn{10}{c}{\textbf{Feature Groups}} \\ \cline{3-8} \cline{10-22} 
 &
   &
  \textbf{L} &
  \textbf{C} &
  \textbf{Rs} &
  \multicolumn{1}{c|}{\textbf{M}} &
  \textbf{G} &
  \multicolumn{1}{c|}{\textbf{D}} &
  \multicolumn{1}{c|}{} &
  \textbf{B} &
  \textbf{F} &
  \multicolumn{1}{c|}{\textbf{T}} &
  \textbf{API} &
  \textbf{REG} &
  \textbf{FILE} &
  \textbf{DIR} &
  \textbf{STR} &
  \textbf{NET} &
  \textbf{SYS} &
  \textbf{DROP} &
  \textbf{SIG} &
  \textbf{Others} \\ \hline
\textbf{EldeRAN}                                                        & $\times$ & $\checkmark$   & $\checkmark$   & $\times$   & $\times$   & $\checkmark$ & $\checkmark$ & $\times$ & $\checkmark$ & $\checkmark$ & $\times$ & $\checkmark$ & $\checkmark$ & $\checkmark$ & $\checkmark$ & $\checkmark$ & $\times$ & $\times$ & $\checkmark$ & $\checkmark$ & $\times$ \\
\textbf{SHIELDFS}                                                       & $\times$ & $\times$   & $\checkmark$   & $\times$   & $\times$   & $\checkmark$ & $\checkmark$ & $\times$ & $\checkmark$ & $\checkmark$ & $\times$ & $\times$ & $\times$ & $\checkmark$ & $\checkmark$ & $\times$ & $\times$ & $\times$ & $\times$ & $\times$ & $\times$ \\
\textbf{ISOT}                                                           & $\times$ & $\checkmark$   & $\checkmark$   & $\times$   & $\checkmark$   & $\checkmark$ & $\times$ & $\times$ & $\checkmark$ & $\checkmark$ & $\times$ & $\checkmark$ & $\checkmark$ & $\checkmark$ & $\checkmark$ & $\checkmark$ & $\times$ & $\checkmark$ & $\checkmark$ & $\checkmark$ & $\checkmark$ \\
\textbf{RanSAP}                                                         & $\times$ & $\times$   & $\checkmark$   & $\checkmark$   & $\times$   & $\checkmark$ & $\times$ & $\checkmark$ & $\checkmark$ & $\checkmark$ & $\times$ & $\times$ & $\times$ & $\checkmark$ & $\checkmark$ & $\times$ & $\times$ & $\times$ & $\times$ & $\times$ & $\checkmark$ \\
\textbf{\begin{tabular}[c]{@{}l@{}}Dynamic \\ Feature\end{tabular}}     & $\times$ & $\checkmark$   & $\checkmark$   & $\checkmark$   & $\times$   & $\checkmark$ & $\times$ & $\checkmark$ & $\checkmark$ & $\checkmark$ & $\times$ & $\checkmark$ & $\checkmark$ & $\checkmark$ & $\checkmark$ & $\times$ & $\checkmark$ & $\times$ & $\times$ & $\times$ & $\checkmark$ \\
\textbf{MarauderMap}                                                    & $\checkmark$ & $\times$   & $\checkmark$   & $\checkmark$   & $\checkmark$   & $\times$ & $\times$ & $\times$ & $\times$ & $\checkmark$ & $\times$ & $\checkmark$ & $\checkmark$ & $\checkmark$ & $\checkmark$ & $\checkmark$ & $\checkmark$ & $\checkmark$ & $\times$ & $\times$ & $\checkmark$ \\
\textbf{MIRAD}                                                          & $\times$ & - & - & - & - & $\times$ & $\times$ & $\times$ & $\checkmark$ & $\times$ & $\times$ & $\checkmark$ & $\checkmark$ & $\checkmark$ & $\checkmark$ & $\times$ & $\times$ & $\times$ & $\times$ & $\times$ & $\times$ \\
\textbf{RCSF}                                                           & $\times$ & $\times$   & $\checkmark$   & $\checkmark$   & $\checkmark$   & $\checkmark$ & $\checkmark$ & $\checkmark$ & $\checkmark$ & $\checkmark$ & $\times$ & $\checkmark$ & $\times$ & $\checkmark$ & $\checkmark$ & $\checkmark$ & $\times$ & $\times$ & $\times$ & $\times$ & $\checkmark$ \\
\textbf{RanSMAP}                                                        & $\times$ & $\times$   & $\checkmark$   & $\checkmark$   & $\times$   & $\checkmark$ & $\times$ & $\checkmark$ & $\checkmark$ & $\checkmark$ & $\times$ & $\times$ & $\times$ & $\checkmark$ & $\checkmark$ & $\times$ & $\times$ & $\times$ & $\times$ & $\times$ & $\checkmark$ \\ \hline

\textbf{\% of $\checkmark$} & $11$  & $33$   & $89$   & $56$   & $33$ & $78$ & $33$ & $44$ & $89$ & $89$ & $0$ & $67$ & $56$ & $100$ & $100$ &  $44$ & $22$ & $22$ & $22$ & $22$ & $67$ \\ \hline \hline

\textbf{\begin{tabular}[c]{@{}l@{}}MLRan - \\ Our Dataset\end{tabular}} & $\checkmark$ & $\checkmark$   & $\checkmark$   & $\checkmark$   & $\checkmark$   & $\checkmark$ & $\checkmark$ & $\checkmark$ & $\checkmark$ & $\checkmark$ & $\checkmark$ & $\checkmark$ & $\checkmark$ & $\checkmark$ & $\checkmark$ & $\checkmark$ & $\checkmark$ & $\checkmark$ & $\checkmark$ & $\checkmark$ & $\checkmark$ \\ \hline

\end{tabular}%
}
\end{table*}



\section{GUIDE-MLRan: Guidelines for developing high-quality ransomware datasets} \label{sec:guide_mlran}
Data quality is a critical determinant of machine learning effectiveness, encompassing both quantitative and qualitative aspects of data evaluation. Its definition varies based on context; for instance, \cite{wang1996beyond} defines data quality as “fit for use,” while \cite{wand1996anchoring} describes it as the alignment between “real-world and system states.” More recently, \cite{zhou2024survey} defined high-quality data as meeting user needs and being purpose-fit for ML tasks, a definition that we adopt in this study. Furthermore, \cite{gong2023survey} emphasises that a high-quality dataset should accurately represent real-world phenomena and be comprehensive and free from biases.

Poor data quality has a profound impact on ML systems. Issues such as mislabeling, inaccuracy, inconsistency, duplication, and overlap can severely impair even the most advanced ML models \citep{chen2021data, zhou2024survey, schwabe2024metric}. In the cybersecurity domain, \cite{tran2022data} empirically demonstrated how data quality impacts ML performance for cyber intrusion detection by evaluating eleven datasets. Their findings underscore the significant influence of data quality on both conventional machine learning models and pre-trained language models, reinforcing the need for rigorous dataset development practices.

While there have been efforts to establish evaluation guidelines for ransomware mitigation solutions, dataset construction has received comparatively little attention. \cite{mcintosh2021ransomware} introduced unified metrics for evaluating ransomware mitigation studies, categorising them into evaluation, output, versatility, and strength. However, their framework does not address how ransomware datasets should be developed, leaving a critical gap in the field. Additionally, while \cite{zhou2024survey} identified eight key data quality dimensions—including completeness, consistency, timeliness, confidentiality, accuracy, standardisation, unbiasedness, and ease of use—these principles were formulated in the context of general ML datasets, with no specific focus on security, malware, or ransomware.

Related work in cybersecurity dataset development has primarily focused on intrusion detection systems (IDSs). \cite{shiravi2012toward} proposed a systematic methodology for generating benchmark IDS datasets, advocating for realistic network traffic capture, accurate labelling, complete interaction records, and comprehensive packet content preservation. Their approach ensures that datasets closely reflect real-world attack conditions, thus improving the reliability of IDS evaluations. Similarly, \cite{tran2022data} proposed nine criteria—reputation, relevance, comprehensiveness, timeliness, variety, accuracy, consistency, duplication, and overlap—to ensure the high quality of IDS datasets.

Although some of these IDS dataset guidelines can be applied to ransomware datasets, they do not account for the unique characteristics of ransomware attacks and the specific requirements of ML-based ransomware detection models. Ransomware exhibits distinct attack behaviours, such as file encryption and system modification, which require a tailored approach to dataset development.

To address this gap, we propose GUIDE-MLRan, a set of guidelines inspired by previous work for developing high-quality behavioural ransomware datasets for ML-based ransomware detection. GUIDE-MLRan is summarised in Table \ref{tab:compliance_main}. The guidelines are based on five key principles: 
\begin{enumerate}
    \item \textit{Sample diversity and representativeness}, ensuring a balanced inclusion of diverse ransomware families and goodware;
    \item \textit{Sample quality and accuracy}, emphasising correct labelling, sufficient dataset size, and samples that cover an extended period;
    \item \textit{Sandbox and testbed requirements}, advocating for realistic execution environments;
    \item \textit{Representative feature extraction}, capturing comprehensive, standardised behavioural features; and
    \item \textit{Documentation, reproducibility, and data extension}, promoting metadata completeness, ethical considerations, public availability, and continuous dataset updates. 
\end{enumerate}
These principles can support the development of robust, reproducible, and well-structured datasets for advancing ransomware detection research. Although these guidelines has been developed specifically for ransomware datasets, they can be applicable to malware datasets in general.

\subsection{Sample diversity and representativeness}

Sample diversity, defined as the inclusion of varied samples, and representativeness, which ensures alignment with real-world distributions, are critical for machine learning datasets \citep{chio2018machine}. \cite{bengio2017deep} highlight that diverse samples enhance model generalisation, enabling them to handle rare and complex scenarios effectively. Similarly, \cite{hastie2017elements} argues that representativeness improves generalisability by ensuring datasets reflect real-world distributions. However, most publicly available ransomware datasets fail to meet these criteria, as collecting diverse and representative samples of ransomware and goodware remains a significant challenge.

A significant issue in constructing ransomware datasets is sample selection bias, where the collected data does not adequately reflect the true distribution of the underlying security problem, which limits model effectiveness \citep{arp2022and, cortes2008sample}. In cybersecurity applications, sampling from the true distribution is inherently difficult and, in some cases, infeasible. While this bias cannot be entirely eliminated, it can be mitigated by ensuring that goodware and ransomware samples are diverse and balanced.

\subsubsection{Diverse ransomware samples}

Ransomware has undergone significant evolution, adapting to technological advancements and security measures over time. Initially, from locker ransomware, which restricts device access until a ransom is paid (e.g., Reveton, Kovter), to crypto-ransomware, which encrypts files and demands payment for decryption, often causing irreversible data loss (e.g., WannaCry, CryptoLocker) \citep{sgandurra2016automated}. The rise of ransomware-as-a-service further expanded its reach by providing ready-made ransomware kits, enabling even non-expert attackers to launch sophisticated attacks (e.g., Hive, GandCrab) \citep{fisher2025vendor}. Modern ransomware integrates RaaS with advanced evasion techniques, such as polymorphism and metamorphism, to bypass traditional detection mechanisms (e.g., Conti, LockBit) \citep{raj2024modern}. Furthermore, modern attacks often include double and triple extortion. Double extortion encrypts data and steals sensitive information, threatening to leak it if the ransom is not paid. Triple extortion adds pressure by targeting third parties or launching DDoS attacks, increasing disruption \citep{kerns2022double, anand2023advancing}.

Capturing this evolution in datasets ensures comprehensive coverage of ransomware behaviours, enhancing machine learning models' ability to detect, classify, and respond to both known and emerging threats \citep{bengio2017deep, hastie2017elements, arp2022and, cortes2008sample, miranda2022debiasing}. However, obtaining ransomware samples can be challenging and downloading them individually can be time-consuming. 


\subsubsection{Diverse goodware samples}

By exposing machine learning models to a broad set of benign applications, diverse goodware samples enhance generalisation, enabling models to distinguish legitimate software from ransomware across various computing environments. Moreover, they prevent bias by ensuring that models do not disproportionately classify certain software categories as malicious  \citep{ashraf2019ransomware}. Additionally, exposure to a wide range of goodware refines detection boundaries, improving the ability to accurately classify known and previously unseen applications \citep{sgandurra2016automated, homayoun2017know, miranda2022debiasing}.

To achieve comprehensive goodware representation, \cite{botacin2021challenges} underscores the importance of curating software that reflects real-world environments, particularly those commonly installed by users within the target system. A well-diversified selection should include applications from multiple categories, such as productivity tools (e.g., Microsoft Office), communication platforms (e.g., Zoom), internet tools (e.g., Firefox), system utilities (e.g., UltraViewer), antivirus software (e.g., Kaspersky Internet Security), lifestyle applications (e.g., Memory-Map), games (e.g., City Car Driving), developer tools (e.g., Python), business software (e.g., Money Manager), and educational programs (e.g., Dictionary). This variety ensures that machine learning models capture normal software behaviours across different domains, reducing the likelihood of misclassification \citep{herrera2023dynamic}.

Furthermore, goodware samples should be sourced from different time periods, incorporating both historical and contemporary software to account for evolving behavioural patterns. This temporal diversity is essential for building realistic datasets, capturing changes in software functionality and execution patterns, and ensuring models remain effective against evolving threats. By encompassing a broad and representative set of goodware, ransomware detection datasets can be used as accurate benchmarks to evaluate the performance and robustness of machine learning models, leading to more reliable and adaptable cybersecurity solutions \citep{miranda2022debiasing}.

However, obtaining diverse goodware samples is challenging due to the vast software landscape, licensing restrictions, and the need for diverse representation. Manually curating such a dataset is time-consuming and requires careful selection to avoid bias while ensuring relevance to real-world environments. 


\subsubsection{Balanced class distribution}

A balanced dataset in machine learning ensures that each class is represented in approximately equal proportions, which prevents model bias and enhances predictive performance. Class imbalance, on the other hand, occurs when one class significantly outnumbers another, leading to biased learning and poor generalisation \citep{miranda2022debiasing}. 

In ransomware detection, maintaining a balanced ratio of ransomware to goodware samples is crucial to prevent the model from favouring one class over the other \citep{onwuegbuche2023enhancing, bengio2017deep}. However, class imbalance in ransomware datasets exists at two levels: inter-class imbalance (between ransomware and goodware) and intra-class imbalance (within ransomware families) \citep{he2009learning}. While achieving inter-class balance is feasible, particularly in dynamic analysis, maintaining intra-class balance is significantly more challenging. This difficulty arises from the natural imbalance in ransomware family prevalence, the rapid evolution of ransomware types, and the scarcity of certain ransomware families \citep{rahman2022limitations, botacin2021challenges}. Some families are widely distributed and frequently analysed, while others remain rare and underrepresented \citep{gong2023survey}.

Despite these challenges, ensuring diversity within ransomware datasets remains essential. A well-balanced dataset should incorporate varied ransomware types and families, reflecting real-world threats while mitigating bias. Achieving a reasonable distribution of ransomware types and families enhances model robustness, enabling machine learning systems to generalise better and improve ransomware detection across different variants.


\subsection{Sample quality and accuracy}
Ensuring sample quality and accuracy is crucial for developing effective machine learning models, particularly in the context of ransomware detection. This requires accurate sample labelling, representativeness and covered time period.

\subsubsection{Accurate labelling}

Accurate labelling is fundamental for supervised learning, ensuring that models are trained on reliable and meaningful data. \cite{gong2023survey} define accuracy as the degree to which data attributes correctly represent the true values of a given concept. In contrast, \cite{cai2015challenges} highlight that accuracy relies on alignment with an agreed “source of truth.” In ransomware detection, labelling accuracy directly impacts detection performance, as incorrect labels can introduce significant bias into the learning process, making systems vulnerable to poisoning attacks and poor detection accuracy \citep{arp2022and,apruzzese2022sok, marcinkowski2024mirad}.

Labelling ransomware samples is particularly challenging due to the lack of readily available ground truth data, creating a chicken-and-egg problem where detection relies on accurate labels, yet obtaining those labels is non-trivial \citep{ceschin2024machine,arp2022and}. The rapid evolution of malware further complicates this process, often requiring manual verification. Manual expert analysis is the most reliable method, but is time-consuming and impractical for large datasets \citep{verma2019data}. To address this challenge, researchers sometimes use VirusTotal Antivirus (AV) majority voting, accepting only malware samples with consistent labels across multiple AV vendors. However, this reduces dataset size and may introduce selection bias, limiting generalisation to ambiguous threats \citep{ceschin2024machine}.

Efforts to standardise malware classification labels are ongoing, with automated labelling tools playing a crucial role. Tools such as Euphony for Android malware \citep{hurier2017euphony} and AVClass for Windows malware \citep{sebastian2016avclass} help streamline classification. A well-labelled dataset should also include detailed metadata, such as ransomware family, type, and sample hashes, enabling nuanced analysis beyond binary classification and reproducibility of results.

One of the common pitfalls in malware research is assuming that all crawled applications from legitimate websites are benign without validation, which can compromise dataset integrity \citep{botacin2021challenges}. To mitigate this problem, researchers should rigorously verify benign samples, for instance, by submitting file hashes to VirusTotal and selecting only those with zero detections across all antivirus engines \citep{pendlebury2019tesseract}. Ensuring high-quality, well-verified labels is crucial for improving model reliability, reducing false positives, and strengthening ransomware detection systems.

\subsubsection{Sample size}

Determining the sufficient number of ransomware and goodware in a dataset to ensure their representativeness is a complex question with no definitive answer. It depends on factors, such as research objectives, the availability of ransomware samples in the wild, the machine learning techniques used, and the level of generalisation required \citep{hirano2022ransap, botacin2021challenges, apruzzese2022sok}. While some methods, such as deep learning, benefit from larger datasets \citep{bengio2017deep}, increasing data volume beyond a certain threshold does not always improve performance. Instead, optimising feature selection and representation strategies may yield better results \citep{ceschin2024machine}. However, representativeness is more critical than size. A small but well-curated dataset that accurately reflects real-world ransomware behaviours is more valuable than a large, biased, or unrepresentative dataset \citep{ceschin2024machine}. Moreover, behavioural ransomware datasets tend to be smaller than static datasets, as behavioural analysis requires individual sample execution and manual validation, making data collection and analysis time-consuming. In summary, the dataset should be large enough to capture common ransomware behaviour patterns while maintaining high-quality, representative samples to enhance model reliability.


\subsubsection{Covered time period}

The time period a dataset covers refers to its suitability for a given task based on its age and longevity \citep{gong2023survey, tran2022data}. \cite{yaseen2023effect} found that models trained exclusively on new ransomware struggle to detect older variants, and vice versa, due to evolving feature sets. However, models trained on temporally diverse datasets achieved 
very high accuracy, demonstrating the importance of incorporating both old and new samples to enhance generalisation and detection robustness.
Since ransomware actors often revive outdated attack methods, it is also crucial to retain historical ransomware samples, such as locker ransomware, alongside modern variants. Similarly, including both recent and older goodware versions improves model generalisation. To maintain temporal consistency and track ransomware evolution, datasets should include timestamps for collected samples \citep{pendlebury2019tesseract}.

\subsection{Sandbox hardening and comprehensive behavioural analysis}
 A sandbox in ransomware analysis is a controlled virtual environment used to execute and observe ransomware behaviour safely without risking harm to the host system or network \citep{sgandurra2016automated, ahmed2024ransomware}. 

\subsubsection{Sandbox hardening}

Sandbox hardening in ransomware behavioural analysis involves strengthening the sandbox environment to resist detection and evasion techniques employed by ransomware \citep{alrawi2024sok}. The objective is to create an analysis environment that closely resembles a real user system, ensuring ransomware executes its full range of malicious behaviours without detecting the sandbox \citep{leguesse2018androneo}. Without effective hardening, many ransomware samples alter their behaviour, delay execution, or terminate when they detect they are running in a sandbox, leading to incomplete or misleading results.

Some malware employs evasion techniques to detect sandbox environments. MITRE ATT\&CK\footnote{MITRE ATT\&CK - Virtualization/Sandbox Evasion: \url{https://attack.mitre.org/techniques/T1497/}} categorises these into three main types:
\begin{enumerate}
    \item System-based checks – Ransomware inspects system attributes to determine if it is running in a sandbox. It checks CPU core count, installed programs, memory, disk size, digital signatures, and system artefacts. If these do not match a real user environment, the malware may refuse to execute.

    \item User activity-based checks – Since sandboxes lack human interactions, ransomware may remain dormant until it detects user activity such as mouse movements, document scrolling, opening applications, or browsing history. Without these actions, the malware assumes it is in a sandbox and delays execution.
    
    \item Time-based evasion – Sandboxes typically execute malware for a limited time, making them vulnerable to delayed execution techniques. Ransomware exploits this by using extended sleep delays, logic bombs, or stalling code, ensuring execution occurs only after the sandbox analysis period expires
\end{enumerate}

Beyond these methods, encryption, obfuscation, and dynamic system alterations further complicate malware detection\footnote{Apriorit: \url{https://www.apriorit.com/dev-blog/545-sandbox-evading-malware}}. \cite{botacin2021challenges} highlight the difficulty of ensuring successful execution in sandbox environments, as some samples fail due to corruption, OS incompatibilities, or evasion tactics. To mitigate this challenge, defining clear execution criteria, such as a minimum number of API calls or observable behaviours, and reporting execution success rates, improves dataset reliability.

To counter sandbox-aware ransomware, hardening strategies should simulate user activity (e.g., opening files, internet browsing), mimic system configurations (e.g., CPU, RAM, OS settings), randomise execution times, and emulate hardware characteristics. These measures reduce detection risk and encourage ransomware to execute fully, capturing more accurate behavioural data.

Execution timeouts are also critical. \cite{kuchler2021does} found that 98\% of executed basic blocks occur within the first two minutes, making a 120-second timeout optimal for capturing malicious behaviour without excessive resource use. This result supports \cite{willems2007toward}, confirming that a two-minute threshold is sufficient for analysing freshly collected malware samples in most cases. Therefore, we implemented a 2-minute (120-second) timeout in this research.
For more sandbox-specific information, we refer the reader to \cite{alrawi2024sok}.

\subsubsection{Comprehensive behavioural analysis} \label{sec:c8}


\cite{gong2023survey} state that a sandbox should collect a comprehensive set of attributes for ransomware behavioural analysis.
These should include application programming interface (API) calls, registry modifications (REG), file operations (FILE), directory activities (DIR), embedded strings (STR), network traffic (NET), system processes (SYS), dropped files (DROP), and signatures (SIG). Missing these attributes can lead to incomplete ransomware profiling, limiting feature extraction for machine learning and detection accuracy.
Therefore, collecting these attributes during dynamic analysis and generating a coherent report is crucial for accurate classification \citep{or2019dynamic}.

\subsection{Representative feature extraction and modelling}
Representative feature extraction ensures that datasets include the most informative attributes, enhancing model accuracy and interoperability. 

\subsubsection{Relevant feature extraction}

After thoroughly analysing ransomware samples and generating a structured report, the next crucial step is extracting comprehensive and relevant features for machine learning-based detection.

\cite{kelleher2020fundamentals} emphasises that selecting meaningful features enhances model performance, efficiency, and interpretability. In ransomware detection, this involves extracting critical behavioural feature groups, such as API calls and registry keys, as outlined in C8 (see Section \ref{sec:c8}), to ensure a detailed and representative characterisation of sample activity.

To automate feature extraction, we develop Python scripts \footnote{MLRan Cuckoo JSON Report Parsers: \url{https://github.com/faithfulco/mlran/tree/main/4_cuckoo_parser_scripts}} that parse Cuckoo Sandbox reports, extracting the nine key behavioural feature groups discussed in C8 (see Section \ref{sec:c8}). These scripts are available publicly to promote reproducibility and standardisation, enabling researchers to streamline feature extraction and enhance ransomware detection frameworks.

\subsubsection{Data preprocessing}

The phrase “garbage in, garbage out” highlights the importance of data preprocessing in machine learning, ensuring that ransomware detection models are trained on clean and well-formatted data \citep{kang2018machine}.

A crucial step in preprocessing is data cleaning, which involves removing duplicates, handling missing values, and standardising feature distributions. Duplicate instances distort evaluation metrics, leading to misleading performance results \citep{chen2021data, zhou2024survey}. Similarly, excessive missing values introduce bias, reducing model reliability \citep{cai2015challenges}. Normalisation and standardisation further ensure consistent feature scales, preventing specific attributes from disproportionately influencing the model.

Beyond cleaning, the dataset format must align with the intended machine learning approach. Sequential data formats (e.g., time-series logs) suit recurrent neural networks (RNNs) for analysing temporal ransomware behaviour, while tabular formats (e.g., CSV, databases) are better suited for classification algorithms like decision trees and logistic regression. Graph-based structures can also help identify ransomware propagation patterns \citep{andronio2015heldroid}.

Feature selection refines the dataset by identifying the most relevant and discriminative attributes, improving model efficiency, interpretability, and generalisation. Approaches include filter methods (statistical relevance), wrapper methods (iterative subset evaluation), and embedded methods (selection during model training). A multi-stage hybrid approach combining these techniques helps eliminate redundancy while retaining essential behavioural indicators \citep{onwuegbuche2023enhancing}.

\subsubsection{Model training and evaluation}

Model training requires careful design choices, bias prevention, and hyperparameter tuning to ensure robust generalisation. A critical issue during training is data snooping bias, which occurs when information from the test set is inadvertently introduced into the training process, leading to overoptimistic performance estimates and poor real-world applicability. \cite{arp2022and} identifies three major types of data snooping bias that must be avoided.

\begin{enumerate}
    \item \textit{Test snooping} occurs when test data is misused for feature selection or hyperparameter tuning, leading to overfitting and misleading performance metrics. To prevent this, training, validation, and test sets must remain strictly separate, with test data used only for final evaluation.
    \item \textit{Temporal snooping} happens when past and future data are mixed during training, ignoring real-world time dependencies. This causes models to perform well on historical data but fail on new ransomware strains. To avoid this, dataset splits should preserve temporal order to reflect ransomware evolution.
    \item \textit{Selective snooping} involves preprocessing based on future knowledge, such as removing outliers using the entire dataset. This creates bias, as such insights would not be available at the time of deployment. Instead, preprocessing should rely only on information accessible during training to ensure fairness and realism.
\end{enumerate}

Beyond avoiding biases, models should be trained on diverse and well-balanced datasets that accurately represent real-world ransomware behaviours. Relying on imbalanced datasets can lead to models that favour the majority class, reducing effectiveness in detecting novel ransomware strains. Additionally, hyperparameter tuning using techniques such as cross-validation helps improve model generalisation while mitigating overfitting.

A robust evaluation framework is essential to assess how well a model generalises to real-world ransomware threats. As \cite{botacin2021challenges} suggest, evaluation criteria should be clearly defined from the start to align with deployment requirements. The evaluation dataset must be representative of the operational environment, avoiding temporal and spatial biases. \cite{pendlebury2019tesseract} highlight that model performance can degrade when tested on data from different time periods or distributions than those used in training. Furthermore, the presence of duplicate samples in the test set can inflate performance metrics, leading to an inaccurate assessment of model effectiveness \citep{tran2022data}.

Choosing the right evaluation metrics is crucial for obtaining a comprehensive performance assessment. While accuracy is often reported, it is insufficient for ransomware detection due to the severe consequences of undetected threats. Instead, evaluation should incorporate Precision, Recall, F1 Score, and False Positive Rate (FPR) to measure detection effectiveness comprehensively. Given the risks associated with undetected ransomware, False Negative Rate (FNR) is particularly important, as a high FNR indicates a failure to detect active ransomware infections \citep{zhou2023malpurifier}. Additionally, class imbalance is common in ransomware datasets, meaning that balanced accuracy and the F1 score should be prioritised over raw accuracy \citep{eren2023semi}. When evaluating models over time, particularly in continual learning settings, performance robustness should be assessed using metrics such as Area Under Time (AUT) to account for concept drift \citep{pendlebury2019tesseract}.

To ensure meaningful comparisons, models should be benchmarked against strong baselines. Using overly complex models without evaluating their performance relative to simpler, well-established approaches can result in unjustified computational overhead. Automated Machine Learning (AutoML) frameworks can assist in identifying competitive baselines, ensuring that performance gains are due to meaningful improvements rather than excessive model complexity \citep{arp2022and}.

Additionally, \cite{rossow2012prudent} emphasises analysing false positives and false negatives to uncover system limitations and improve detection robustness, as error rates alone offer limited insight. In our work, we perform a thorough misclassification analysis using explainable AI methods. 

\subsection{Documentation, reproducibility, and data extension}

Comprehensive documentation, reproducibility, and regular data updates are essential for creating reliable ransomware datasets. Detailed metadata and transparent processes ensure that datasets are understandable, reusable, and adaptable. These practices enhance collaboration, validate results, and maintain relevance in addressing evolving ransomware threats. \cite{botacin2021challenges} highlights that using non-reproducible methodologies for dataset definition and experimental design is a significant pitfall in malware research. Reproducibility is essential for scientific integrity as it enables other researchers to verify or challenge results, thereby advancing the field.

\subsubsection{Availability of contextual and metadata information}

Providing comprehensive contextual and metadata information about samples, sandboxes, datasets, and methods is fundamental to ensuring transparency, interpretability, and reproducibility in ransomware detection research \citep{gong2023survey, rossow2012prudent}. Without detailed documentation, experimental findings become difficult to validate, compare, or replicate, hindering progress and limiting the applicability of research \citep{mcintosh2021ransomware, holzinger2019machine}.

Metadata at the sample level must capture essential attributes such as sample type, source, hash, description, timestamp, and classification (ransomware family and type or goodware category). These details provide critical context for assessing dataset representativeness, preventing biases such as temporal snooping, and ensuring robust experimental design \citep{cai2015challenges}. For instance, to avoid temporal snooping, it is critical for the timestamp of each sample to be provided \citep{arp2022and}. 

At the dataset level, documentation should include collection methods, preprocessing steps, feature extraction techniques, and class distribution. This information is crucial for reconstructing datasets, adapting them to new research objectives, and ensuring fair comparisons between detection models \citep{herrera2023dynamic}. Without clear documentation, differences in data preprocessing—such as handling of missing values or selection of dynamic behavioural features—can lead to inconsistent model performance and misleading conclusions.

Equally important is documenting the sandbox or testbed settings used for sample analysis. Sandboxes differ in their configurations, execution parameters, and evasion countermeasures, which can significantly impact how ransomware behaves during execution. Therefore, system details such as OS version, installed software, hardware specifications, sandbox type (e.g., Cuckoo, BitVisor), execution timeouts, network connectivity settings, and anti-evasion mechanisms must be explicitly recorded \citep{rossow2012prudent}. Inadequate sandbox documentation compromises result reproducibility and makes it difficult to assess detection effectiveness across different environments \citep{alrawi2024sok, leguesse2018androneo}.

Beyond the dataset and sandbox documentation, methodological transparency is essential. Reporting hyperparameter settings, feature selection strategies, training-validation splits, and evaluation metrics ensures that model comparisons are fair and reproducible. \cite{cai2015challenges} highlight that poorly documented methods introduce ambiguity, making it difficult to distinguish whether performance variations arise from the model’s capability or differences in data handling.

\subsubsection{Public availability of data and code}

Open access allows independent validation, facilitates model comparisons, and drives innovation \citep{praveen2023current}.
High-quality datasets should be hosted on stable repositories (e.g., Zenodo, IEEE DataPort) with version control and open-access licensing (e.g., Creative Commons or GNU GPL) to ensure long-term availability and broad adoption \citep{zhou2024survey}. Likewise, publishing code for data processing, model training, and evaluation prevents inconsistencies in feature extraction and preprocessing, ensuring results are replicable \citep{arp2022and}. Publicly accessible code allows other researchers to validate results, refine methodologies, and build upon existing work, contributing to the collective advancement of the field.

\subsubsection{Ethical and legal considerations}

Ensuring ethical and legal compliance when curating and sharing ransomware datasets is essential to prevent misuse, protect privacy, and adhere to regulatory standards. Researchers must balance open data access with responsible handling to maintain transparency while mitigating risks associated with malware distribution and sensitive data exposure \citep{thomas2017ethical}. Datasets should comply with cybersecurity laws, institutional guidelines, and ethical frameworks, such as the General Data Protection Regulation (GDPR) in the EU and the Computer Fraud and Abuse Act (CFAA) in the USA. Adhering to third-party data usage policies and licensing agreements, such as those from VirusTotal, is also critical to ensure regulatory compliance \citep{nasir2024ethical}.

To prevent misuse, ransomware source code should not be publicly released. Instead, researchers should share derived features, metadata, and execution logs, which provide meaningful insights while minimising security risks \citep{thomas2017ethical}. If access to ransomware binaries is necessary, controlled distribution mechanisms—such as request-based access for vetted researchers—should be implemented to prevent unauthorised use. Additionally, datasets must be anonymised to remove personally identifiable information (PII), including usernames, file paths, IP addresses, and domain names, to prevent privacy breaches.

Ensuring security and containment when handling live ransomware samples is essential to prevent unintended harm. Ransomware analysis systems must be fully isolated from real systems and networks, following institutional security policies \citep{alrawi2024sok}. Implementing containment measures helps redirect malicious traffic and mitigate risks, making it a key ethical consideration. Additionally, any security breaches within the containment environment should be monitored and documented to improve system resilience and enhance future threat mitigation strategies \citep{rossow2012prudent}.

Ethical considerations also require clear documentation outlining the dataset’s intended use, risks, and restrictions. Responsible disclosure ensures that datasets support cybersecurity research without enabling malicious activities. Researchers should consult ethical review boards or cybersecurity committees when handling datasets that pose dual-use risks, meaning they could be exploited for harmful purposes \citep{gao2024documenting}. Moreover, appropriate licensing (e.g., Creative Commons or GNU GPL) should define terms of use, ensuring responsible data sharing while acknowledging contributors.

By following these guidelines, researchers can curate and share ransomware datasets responsibly, promoting compliance, security, and collaboration while ensuring that datasets remain valuable for advancing ransomware detection research.

\subsubsection{Continuous updates}
Given the rapid evolution of ransomware, datasets must be continuously updated to capture new variants and emerging behavioural patterns. Regular updates enhance dataset relevance, ensuring machine learning models remain effective against evolving threats \citep{herrera2023dynamic}. By openly releasing our dataset and code, we enable researchers to extend, refine, and adapt the dataset, fostering collaborative advancements in ransomware detection as attack strategies evolve.

\section{The MLRan dataset} 
\label{sec:mlran_dataset}
This section describes the MLRan dataset, the methods used for sample collection and analysis, and the feature extraction techniques we employed to run the ML experiments. The compliance of the MLRan dataset with the GUIDE-MLRan guidelines is summarised in Table~\ref{tab:compliance_main} and further elaborated in this section.

\subsection{Dataset description}




\begin{figure}[htpb]
\centering 

\includegraphics[width=0.99\columnwidth]{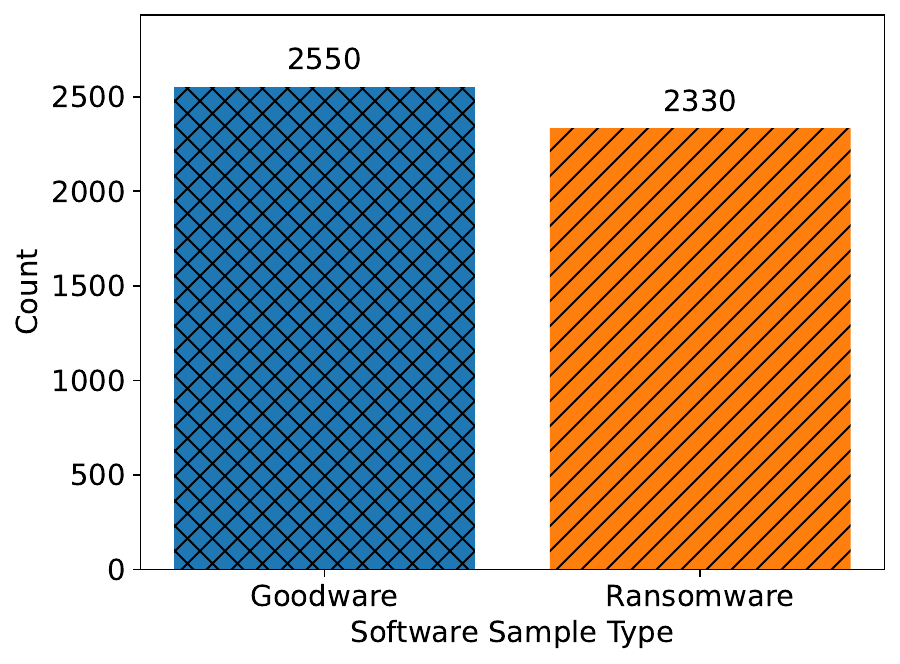}%
\caption{Distribution of software sample types in the MLRan dataset. The dataset contains a total of 4880 samples, split into 2550 (52.25\%) Goodware and 2330 (47.75\%) Ransomware. The dataset is relatively balanced, with only a slight difference between the two categories.}
\label{fig:sample_type}
\end{figure}

Figure \ref{fig:sample_type} illustrates the distribution of software sample types within the MLRan dataset. The dataset contains a total of 4880 samples, with 2550 (52.25\%) representing Goodware and 2330 (47.75\%) corresponding to Ransomware. This shows a relatively even distribution between the two categories, with only a small, negligible proportion difference. Such a balanced dataset is advantageous for training machine learning models, as it reduces the risk of class imbalance affecting model performance. This balance ensures that the model can effectively differentiate between Goodware and Ransomware without being biased towards one class.

\begin{figure}[htpb]
\centering 
  \includegraphics[width=0.99\columnwidth]{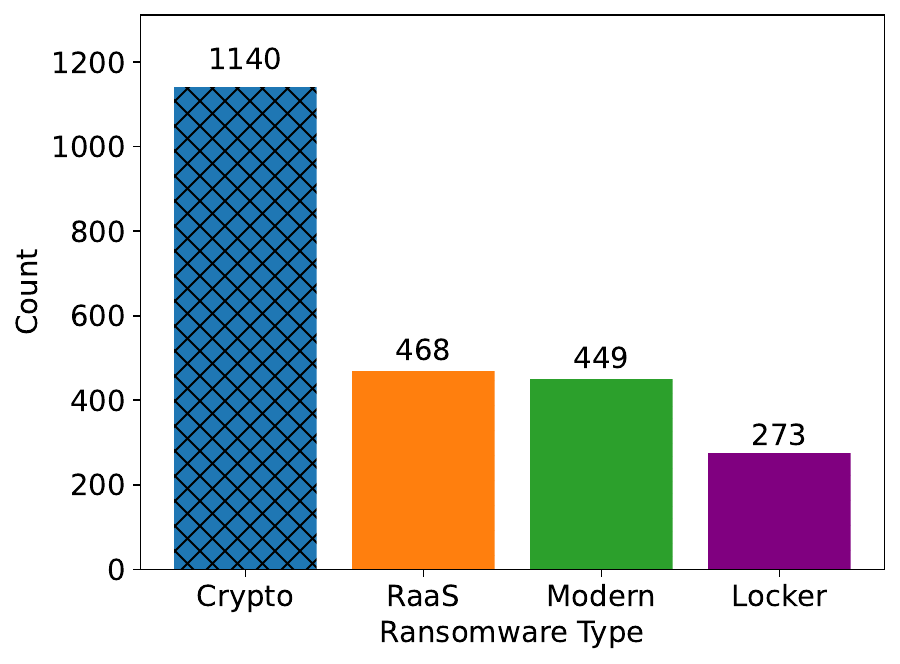}%
\caption{Distribution of ransomware types in the MLRan dataset. The dataset contains a total of 2330 Ransomware samples, split into 1140 (48.92\%) Crypto, 468 (20.08\%) RaaS, 449 (19.27\%) Modern, and 273 (11.72\%) Locker. }
\label{fig:ransomware_type}
\end{figure}

\begin{figure*}[htpb]
    \centering
    \includegraphics[width=1\textwidth]{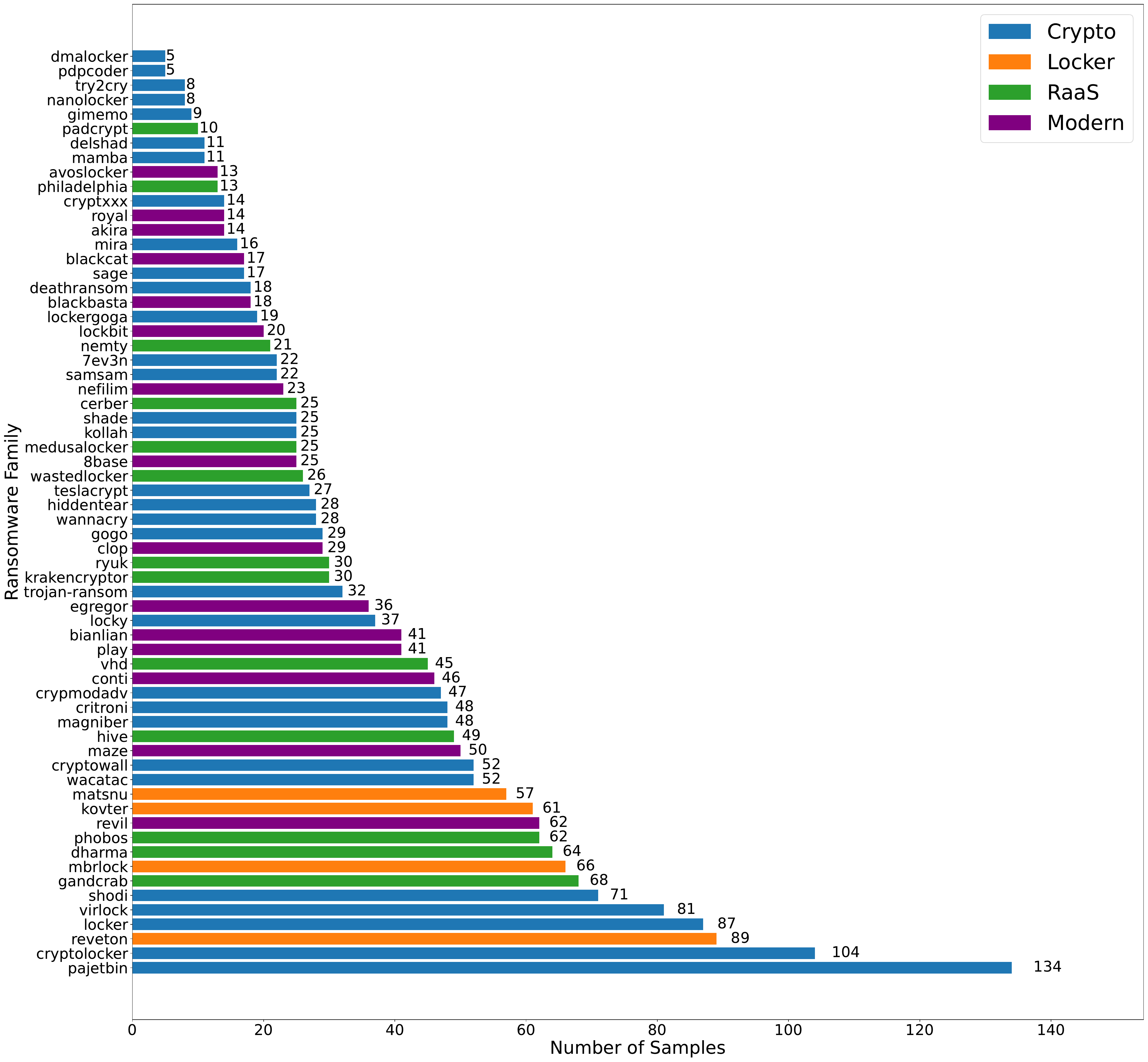}
    \caption{Distribution of ransomware families, colour-coded by their respective ransomware types. The numbers on the bars represent the number of samples from each ransomware family found in the MLRan Dataset. The dataset includes a total of 64 ransomware families, classified into four categories, as shown in the legend and colour-coded in the bars: 32 families belong to the Crypto type, 15 are Modern, 13 are RaaS, and 4 are Locker.}
    \label{fig:ransomware_family}
\end{figure*}

Figure \ref{fig:ransomware_type} shows the distribution of ransomware types in the MLRan dataset. Out of a total of 2330 ransomware samples, 48.92\% (1140 samples) belong to the Crypto type, making it the most prevalent type in the dataset. The next most common types are RaaS with 20.08\% (468 samples), followed by Modern at 19.27\% (449 samples), and finally, Locker, which accounts for 11.72\% (273 samples). A similar distribution pattern is observed at the family level, as illustrated in Figure \ref{fig:ransomware_family}. The dataset contains a total of 64 ransomware families. Specifically, 32 families belong to the Crypto type, 15 are categorised as Modern, 13 fall under RaaS, and 4 correspond to the Locker type. 
This distribution satisfies the criteria (C1) of including diverse ransomware samples.

The distributions of ransomware types in Figure \ref{fig:ransomware_type} and ransomware families in Figure \ref{fig:ransomware_family} in the MLRan dataset reflect ransomware's current and historic distribution in the wild. 
While locker ransomware, which locks users out of their systems without encrypting files, still poses a threat, it is currently less prevalent than other types. Historically, crypto ransomware has been the most prominent, with several high-profile families making headlines. For instance, according to the UK's \cite{ncsc2018cyber} report, the WannaCry attack in 2017 infected over 300,000 computers across 150 countries, including the UK's National Health Service (NHS). 

Crypto ransomware remains a significant threat. Emerging insights from the VirusTotal 2021 report\footnote{VirusTotal report on Ransomware in the Global Context: \url{https://blog.virustotal.com/2021/10/ransomware-in-global-context.html}} reveal that 95\% of all ransomware samples were Windows-based executable files or dynamic link libraries, typically associated with this type. However, the landscape has shifted towards RaaS and more sophisticated modern ransomware variants. RaaS has seen a dramatic rise in recent years, contributing significantly to the increase in ransomware attacks. Notable RaaS families include Ryuk, which was responsible for one-third of all ransomware attacks in 2020, and Cerber, one of the earliest RaaS offerings \footnote{Security Magazine: \url{https://www.securitymagazine.com/articles/93769-ryuk-ransomware-responsible-for-one-third-of-all-ransomware-attacks-in-2020}}. 

Modern ransomware attacks often combine multiple techniques and have become increasingly sophisticated. According to Statista\footnote{Statista: \url{https://www.statista.com/statistics/1475291/most-detected-ransomware-types-worldwide/}}, prominent families such as LockBit, Akira, BlackCat, Play, and Royal are the top five most-detected ransomware attacks in 2023, which are all included in our dataset, as illustrated in Figure \ref{fig:ransomware_family}. These modern variants often incorporate features from crypto and RaaS ransomware types while adding new tactics such as double or triple extortion, which involves not only encrypting data but also threatening to leak sensitive information obtained through data exfiltration or launch DDoS attacks\footnote{Sealpath: \url{https://www.sealpath.com/blog/ransomware-raas-operations-guide/}}.

\begin{figure}[htpb]
\centering 
  \includegraphics[width=0.99\columnwidth]{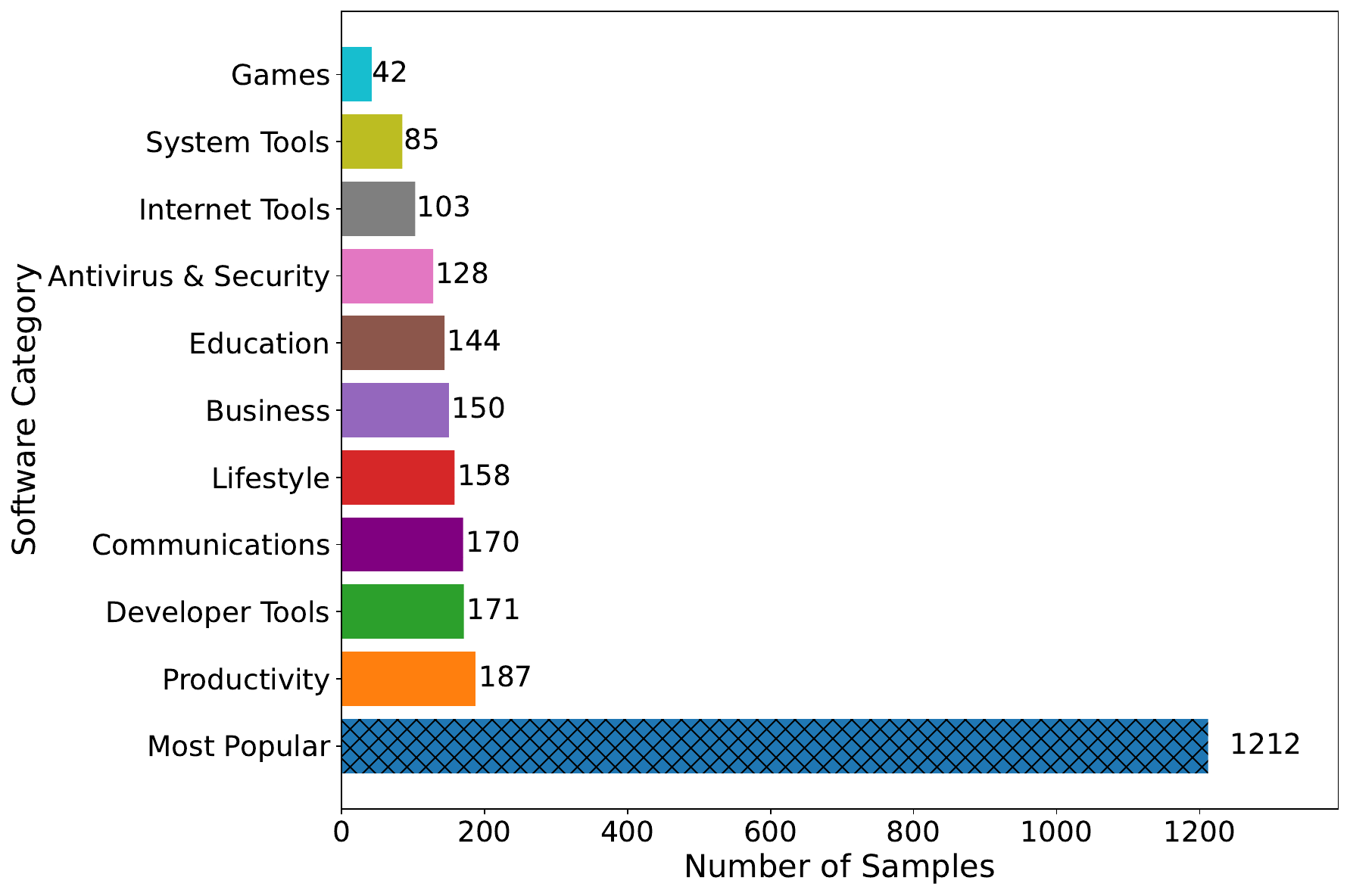}%
\caption{Distribution of goodware sample categories in the MLRan Dataset. The goodware samples contain 11 categories, with the Most Popular category having the highest sample count of 1212 samples, representing 47.53\% of the total. The Productivity category follows with 187 samples, accounting for 7.33\%. The exact number of samples for each category is displayed on the corresponding bars.}
\label{fig:goodware_category}
\end{figure}

Furthermore, the MLRan dataset encompasses various goodware samples, as illustrated in Figure \ref{fig:goodware_category}. These samples span 11 broad categories of legitimate applications, comprising a total of 2,550 samples. The most commonly used applications by individuals constitute the largest proportion, accounting for 47.52\% (1,212 samples). In addition, the dataset includes applications from 10 other categories, ensuring a comprehensive representation of goodware. Specifically, productivity applications account for 7.33\% (187 samples), followed by developer tools at 6.71\% (171 samples), communications software at 6.67\% (170 samples), and lifestyle applications at 6.20\% (158 samples). Moreover, business-related applications represent 5.88\% (150 samples), while antivirus and security software comprise 5.02\% (128 samples). Other categories include internet tools (4.04\%, 103 samples), system tools (3.33\%, 85 samples), and games, which form the smallest category at 1.65\% (42 samples). 

The inclusion of these diverse, balanced goodware and ransomware samples ensures that MLRan satisfies the guidelines for sample diversity and representativeness (C1, C2 and C3).

\begin{figure*}[htpb]
    \centering
    \includegraphics[width=1\textwidth]{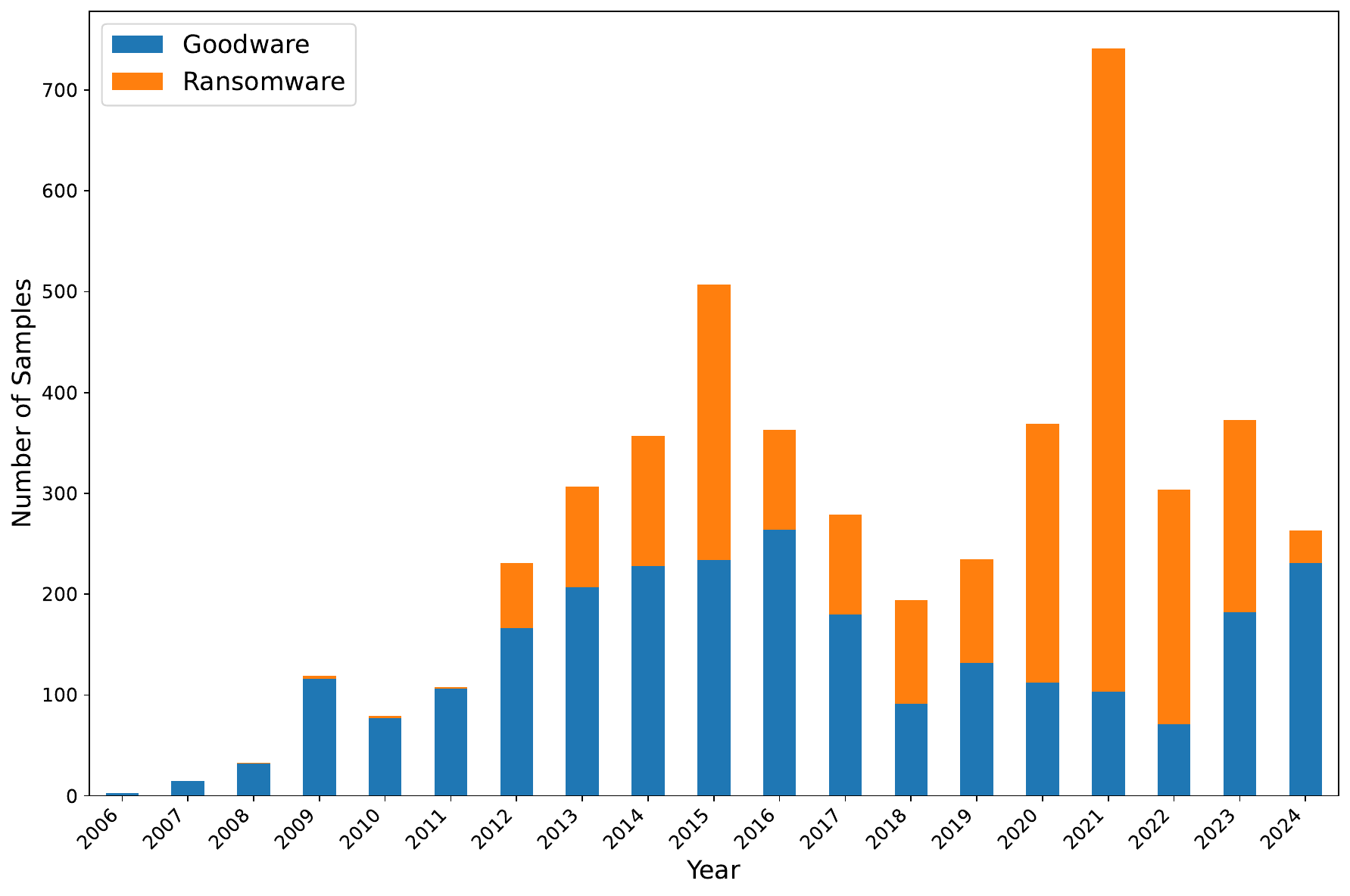}
    \caption[The stacked bar chart illustrates the distribution of samples in the MLRan dataset across different years, based on their first submission to VirusTotal, segmented by sample type (Goodware and Ransomware). The first submission timestamp from VirusTotal was used as it provides more reliable temporal information than creation dates, which malware authors often manipulate. Between 2006 and 2011, the dataset predominantly consisted of Goodware samples. Starting in 2012, a steady increase in Ransomware samples is observed, while Goodware samples remained relatively stable, except for a notable spike in Ransomware during 2020 and particularly in 2021, likely driven by the surge in cybercrime during the COVID-19 pandemic.]%
    {The stacked bar chart illustrates the distribution of samples in the MLRan dataset across different years, based on their first submission to VirusTotal, segmented by sample type (Goodware and Ransomware). The first submission timestamp from VirusTotal was used as it provides more reliable temporal information compared to creation dates, which are often manipulated by malware authors\footnotemark. Between 2006 and 2011, the dataset predominantly consisted of Goodware samples. Starting in 2012, a steady increase in Ransomware samples was observed, while Goodware samples remained relatively stable, except for a notable spike in Ransomware during 2020 and particularly in 2021, likely driven by the surge in cybercrime during the COVID-19 pandemic.}
    \label{fig:timestamp}
\end{figure*}

To demonstrate that the dataset meets the timeliness criterion (C6), we present a stacked bar chart in Figure \ref{fig:timestamp}, illustrating the distribution of ransomware and goodware samples across different years. The plot reveals that ransomware samples in the MLRan dataset span from 2008 to 2024, while goodware samples cover a slightly broader period from 2006 to 2024.

Between 2006 and 2011, the dataset predominantly comprises goodware samples, reflecting the earlier prevalence of benign software. However, from 2012 onwards, a steady increase in ransomware samples was observed, whereas the number of goodware samples remained relatively stable. Notably, a significant surge in ransomware samples occurred in 2020, with an even sharper increase in 2021. This spike is likely attributed to the escalation of cybercrime during the COVID-19 pandemic, which saw a rise in ransomware attacks targeting organisations and individuals worldwide \citep{baig2023ransomware, minnaar2021cyberattacks}. The timestamps for the samples were determined by their first submission to VirusTotal, providing a more reliable temporal reference. This method is more robust than relying on creation dates, which malware authors frequently manipulate to evade detection \citep{jiang2024benchmfc}.

Furthermore, additional contextual and metadata information satisfying criterion (C12), including sample hashes, descriptions, and other relevant attributes, is available in the corresponding CSV files within our repository\footnote{MLRan Metadata: \url{https://github.com/faithfulco/mlran/tree/main/2_collected_samples_metadata}}.

\footnotetext{VirusTotal Files Info.: \url{https://docs.virustotal.com/reference/files}}

\subsection{Sample collection}



\begin{figure}[htpb]
\centering 
  \includegraphics[width=1\columnwidth]{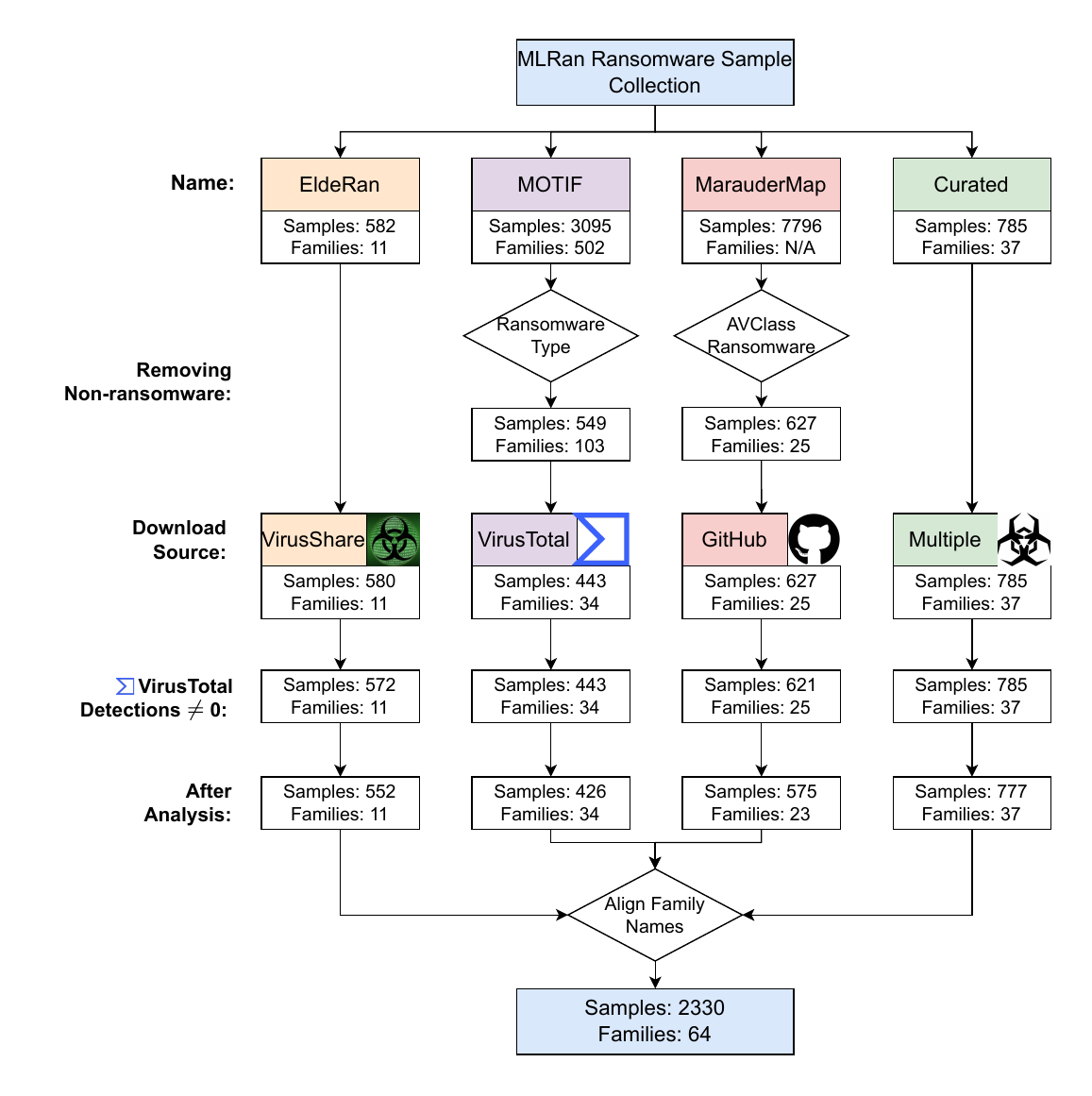}%
\caption{MLRan ransomware sample collection methodology. The diagram illustrates the four primary sources from which ransomware samples were obtained for the MLRan dataset: EldeRan (552 samples), MOTIF (426 samples), MarauderMap (575 samples), and a curated collection (777 samples). }
\label{fig:ransomware_design}
\end{figure}

Figure \ref{fig:ransomware_design} illustrates the methodology used for ransomware sample collection, outlining the steps taken to ensure the ransomware samples' diversity, representativeness, uniqueness, and authenticity. The dataset comprises samples from four primary sources, each subjected to rigorous validation procedures. We developed custom scripts to automate the downloading of both ransomware and goodware samples. These scripts are publicly available and properly documented in our repository\footnote{MLRan Sample Collection Scripts: \url{https://github.com/faithfulco/mlran/tree/main/1_sample_collection_scripts}}.

The first source is Elderan \cite{sgandurra2016automated}, a dataset published in 2016 containing 582 ransomware samples with their respective names obtained by manually clustering each ransomware into a well-established family name. We downloaded these samples from VirusShare using a custom-designed script. We verified their hashes against VirusTotal to confirm their authenticity, ensuring each sample had a non-zero detection score. Additionally, duplicate samples were identified and removed to maintain dataset integrity.

The second source is MOTIF \cite{joyce2023motif}, a malware reference dataset with ground truth family labels. MOTIF addresses the issue of noisy labels in malware analysis by curating samples based on hundreds of open-source threat intelligence reports published by reputable cybersecurity organisations. From the 3,095 samples spanning 502 malware families, we filtered those classified as ransomware, yielding 549 samples across 103 families. Using their hashes, we attempted to retrieve the corresponding binaries from VirusTotal, successfully obtaining 443 samples across 34 families. After duplicate removal and ensuring that VirusTotal detections remained non-zero, the sample count remained unchanged. However, dynamic analysis conducted using the Cuckoo sandbox successfully processed 426 ransomware samples spanning 34 distinct families.

The third source is MarauderMap \cite{hou2024empirical}, which contains 7,796 malicious samples but lacks family labels. To categorise these samples, we applied AVClass \cite{sebastian2016avclass} for automatic family classification. We then conducted a manual verification process to ensure that only ransomware samples were retained, resulting in 627 samples across 25 ransomware families. We downloaded these samples from GitHub, and after removing duplicates and ensuring a non-zero VirusTotal detection score, we obtained 621 unique samples. Following analysis in the Cuckoo sandbox, 575 ransomware samples were successfully processed without errors, representing 25 distinct families.

The fourth source consists of ransomware samples curated by the authors, following a methodology similar to that of \cite{joyce2023motif}. We collected ransomware sample hashes from open-source threat intelligence reports containing detailed ransomware analyses published by reputable cybersecurity organisations. Additionally, we sourced samples from various malware repositories, including MalwareBazaar, VirusShare, VirusTotal, and Hybrid-Analysis. The sample size remained unchanged after removing duplicates and verifying non-zero VirusTotal detections. However, dynamic analysis using the Cuckoo sandbox yielded successful results for only 777 ransomware samples, covering 37 unique families.

Finally, we standardised family names to ensure consistency across sources, accounting for aliasing and duplicate families that appeared in multiple sources. This alignment process improved dataset integrity and ensured uniform classification of ransomware families.

\begin{figure}[htpb]
\centering 
  \includegraphics[width=0.5\columnwidth]{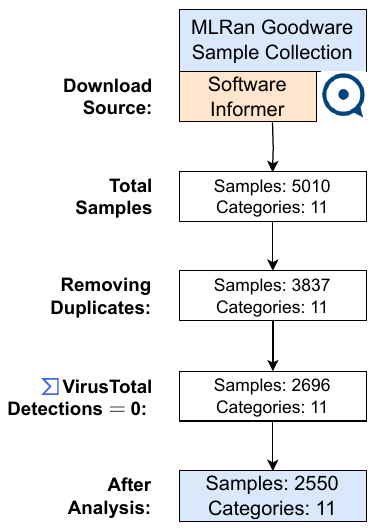}%
\caption{MLRan goodware sample collection methodology. All the goodware samples were downloaded from the Software Informer website and they cut across 11 different categories.}
\label{fig:goodware_design}
\end{figure}

Figure \ref{fig:goodware_design} illustrates the methodology employed for goodware sample collection, detailing the steps taken to ensure diversity, representativeness, uniqueness, and authenticity. We retrieved all goodware samples from Software Informer using a custom-designed script we developed, resulting in an initial dataset of 5,010 samples spanning 11 categories. Following the removal of duplicate entries, the dataset included 3,837 unique samples.

As highlighted by \cite{botacin2021challenges}, a common pitfall in malware research is the assumption that crawled applications are inherently benign without proper validation. To mitigate this issue, we verified the legitimacy of the goodware samples by submitting their hashes to VirusTotal for analysis. Only those samples that received zero detections across all antivirus engines were retained, yielding a refined dataset of 2,696 goodware samples. Finally, dynamic analysis using the Cuckoo sandbox successfully processed 2,550 samples, ensuring coverage across all 11 categories.

\subsection{Sample analysis}


To ensure accurate and reliable behavioural analysis of ransomware and benign software, we employed \textbf{Cuckoo Sandbox}, an open-source automated malware analysis system. Cuckoo allows executing suspicious files in an isolated environment, monitoring their behaviour and generating detailed reports \citep{liu2014poster}. This capability is essential for ransomware analysis, as it enables the observation of encryption activities, network communications, and persistence mechanisms in a controlled setting. Given its adaptability, effectiveness and popularity, Cuckoo was selected as the primary analysis tool.

We configured the sandbox environment to mimic a real-world system closely. We installed VMware Workstation on a Windows host machine and deployed Ubuntu 18.04 as the base operating system. We installed Cuckoo Sandbox v2.0.7 alongside a Windows 7 virtual machine, which served as the execution environment for ransomware and goodware samples. This configuration balanced compatibility with modern malware while ensuring a stable, reproducible analysis environment.

Since sandbox-aware malware can detect and evade execution within virtualised environments \citep{liu2014poster}, we applied several countermeasures. To enhance realism, we allocated four CPUs, 4GB of RAM, and sufficient disk space to the virtual machine. Additionally, we installed commonly used software, including Adobe PDF Reader, .NET Framework, Java, Flash, Visual C++ Redistributable (vcredist 2015u3), and Internet Explorer 11, using VMCloak to automate installation. This ensured the virtual machine resembled a genuine user system.

To further prevent detection, we integrated Cuckoo’s \textit{Human Auxiliary Module}\footnote{Cuckoo Human Module: \url{https://github.com/cuckoosandbox/cuckoo/blob/master/cuckoo/data/analyzer/windows/modules/auxiliary/human.py}}, which simulates user interactions such as keyboard input and mouse movements. The module randomises cursor movements and speeds to mimic human behaviour, reducing the likelihood of detection by sandbox-aware malware. 

Another auxiliary module we integrated into the analysis environment is \textit{Disguise Auxiliary Module}\footnote{Cuckoo Disguise Module: \url{https://github.com/cuckoosandbox/cuckoo/blob/master/cuckoo/data/analyzer/windows/modules/auxiliary/disguise.py}}, which enhances realism by modifying system attributes to evade malware detection. Specifically, the module randomises the Windows Product ID, modifies SCSI identifiers to replace virtual machine-related strings with realistic hardware names, and alters BIOS information, including system and video BIOS dates and versions. Additionally, it patches ACPI tables to replace VirtualBox-related identifiers with manufacturer names, modifies processor details to appear as physical hardware, and changes manufacturer and product names in system information. It also updates hard drive paths to resemble real hardware configurations. Additionally, to counteract time-based evasion techniques, where ransomware delays execution to avoid detection, we set an analysis timeout of 120 seconds to capture delayed malicious behaviour. These modifications prevent ransomware from identifying the environment as a sandbox, further improving the reliability of behavioural analysis. 

\subsection{Feature extraction}


As summarised in Table \ref{tab:feature_category_summary}, the feature extraction methodology in this study focuses on nine categories derived from Cuckoo Sandbox reports, i.e. API calls, registry key operations, file operations, directory operations, extracted strings, network activity, system resource usage, dropped files extensions and types, and behavioural signatures. These features capture key behavioural aspects of software and are represented using a \textbf{binary presence approach}, where each feature is assigned a value of 1 if present or 0 if absent. This method aligns with prior research~\citep{sgandurra2016automated}, ensuring consistency in malware analysis.


We followed the approach in Algorithm \ref{alg:process_reports} to extract API calls from Cuckoo Sandbox JSON reports. The algorithm loads each report (Lines 4–5), extracts the apistats section (Line 6), and creates a dictionary where each observed API call is recorded with a binary indicator prefixed by API: (Lines 7–10). Sample IDs are extracted from filenames and used to index the data (Line 11). All records are stored and converted into a DataFrame (Line 13), with missing values filled as zeros (Line 14) and sorted by sample ID (Line 15). The final DataFrame represents each sample as a vector of API call presences (Line 16).


\begin{table*}[htpb]
    \centering
    \caption{Summary of extracted feature categories from Cuckoo Sandbox reports. Each category represents a key behavioural aspect of the analysed software. The table details the feature descriptions, their extraction sources within the Cuckoo report, and example feature names.}
    \label{tab:feature_category_summary}
    \renewcommand{\arraystretch}{1.3}
    \resizebox{\textwidth}{!}{%
    \begin{tabular}{|l|p{5cm}|p{4cm}|p{5cm}|}
        \hline
        \textbf{Feature Category} & \textbf{Description} & \textbf{Cuckoo Report Section} & \textbf{Example Features Names} \\ 
        \hline
        \textbf{API Calls (API)} & Captures system interactions related to file manipulation, memory access, network activity, and process execution. These calls indicate how a program interacts with the OS. & \texttt{behavior → apistats} & \texttt{\detokenize{API:GetAdaptersInfo}}, \texttt{\detokenize{API:CreateProcessInternalW}}, \texttt{\detokenize{API:CryptUnprotectData}}, \texttt{\detokenize{API:DnsQuery_UTF8}} \\ 
        \hline
        \textbf{Registry Keys (REG)} & Tracks registry modifications, including keys opened, deleted, read, or written. These modifications are often used for persistence and disabling security features. & \texttt{behavior → summary → regkey\_opened, regkey\_deleted, regkey\_read, regkey\_written} & \texttt{\detokenize{REG:DELETED:HKEY_CLASSES_ROOT\*\shell\Secure Eraser}}, \texttt{\detokenize{REG:WRITTEN:\REGISTRY\USER\.DEFAULT\SOFTWARE\Piriform\Recuva\Language}} \\ 
        \hline
        \textbf{File Operations (FILE)} & Monitors file access patterns, including creation, deletion, opening, and writing, which can be indicative of ransomware encryption attempts. & \texttt{behavior → summary → file\_created, file\_deleted, file\_written, file\_opened} & \texttt{\detokenize{FILE:CREATED:c:\!!!how_to_decrypt!!! .txt}}, \texttt{\detokenize{FILE:WRITTEN:z:\boot\recovery_instructions.html}} \\ 
        \hline
        \textbf{Directory Operations (DIR)} & Identifies directory creation and enumeration, which ransomware often uses to access and encrypt multiple files. & \texttt{behavior → summary → directory\_created, directory\_enumerated} & \texttt{\detokenize{DIRECTORY:CREATED:\.\c:\programdata}}, \texttt{\detokenize{DIRECTORY:ENUMERATED:z:\boot\sv-se\*}} \\ 
        \hline
        \textbf{Strings (STR) } & Extracts human-readable text from binaries, including error messages, function names, registry paths, and command-line arguments, providing contextual clues to software intent. & \texttt{strings} & \texttt{\detokenize{STRING:snd_clipcopy}}, \texttt{\detokenize{STRING:m\device\harddiskvolume2\program files\mcafee\engine\avvclean.datp}} \\ 
        \hline
        \textbf{Network Activity (NET) } & Tracks network communication, such as IP connections, domain resolutions, and host interactions, which are crucial for identifying C2 communications. & \texttt{network → connects\_ip, connects\_host, resolves\_host} & \texttt{\detokenize{NETWORK:CONNECTS_HOST:104.131.182.103}}, \texttt{\detokenize{NETWORK:RESOLVES_HOST:yahoo.com}} \\ 
        \hline
        \textbf{System Resources (SYS) } & Examines how software interacts with system resources, such as DLLs, command-line execution, mutex creation, and GUID usage. & \texttt{behavior → summary → dll\_loaded, command\_line, mutex, guid} & \texttt{\detokenize{SYSTEM:DLL_LOADED:user32.dll}}, \texttt{\detokenize{SYSTEM:MUTEX:$Mutex_XYZ$}} \\ 
        \hline
        
        \textbf{\begin{tabular}[c]{@{}l@{}}Dropped Files \\ Extensions and \\ Types (DROP)\\ \end{tabular}} & Captures files created or modified by the sample, including file types and extensions, to detect ransomware payloads or encrypted file formats. & \texttt{dropped} & \texttt{\detokenize{DROP:EXTENSION:.exe}}, \texttt{\detokenize{DROP:TYPE:zip_archive_data}} \\ 
        \hline
        \textbf{Signatures (SIG) } & Detects predefined behavioural patterns associated with malware, including anti-VM detection and ransomware indicators. & \texttt{signatures} & \texttt{\detokenize{SIGNATURE:allocates_execute_remote_ process}}, \texttt{\detokenize{SIGNATURE:antiemu_wine}} \\ 
        \hline
    \end{tabular}
    }%
\end{table*}

\begin{algorithm}[htpb]
\caption{Processing Cuckoo Sandbox JSON reports for API calls}
\label{alg:process_reports}
\SetKwInOut{Input}{Input}
\SetKwInOut{Output}{Output}

\Input{Folder \texttt{reports\_folder} containing Cuckoo JSON reports}
\Output{DataFrame \texttt{df} summarizing API calls per sample}

\textbf{Initialize} empty list \texttt{data} to store API call summaries\;
\textbf{Initialize} empty list \texttt{file\_ids} to store report IDs\;
\textbf{Retrieve} all filenames in \texttt{reports\_folder} matching digits or ending with \texttt{.json}\;

\ForEach{\texttt{filename} in \texttt{all\_files}}{
    \textbf{Load} JSON file \texttt{filename} into \texttt{report}\;
    \textbf{Extract} \texttt{apistats} section from \texttt{report.behavior.apistats}\;
    
    \textbf{Initialize} empty dictionary \texttt{summary} for API calls\;
    \ForEach{\texttt{api\_dict} in \texttt{apistats.values()}}{
        \ForEach{\texttt{api} in \texttt{api\_dict.keys()}}{
            \texttt{summary}[``API:\texttt{api}''] $\gets 1$\;
        }
    }
    
    \textbf{Extract} \texttt{file\_id} from \texttt{filename} (removing extension if necessary)\;
    \textbf{Append} \texttt{summary} to \texttt{data} and \texttt{file\_id} to \texttt{file\_ids}\;
}

\textbf{Create} DataFrame \texttt{df} from \texttt{data}, indexed by \texttt{file\_ids}\;
\textbf{Fill} missing values in \texttt{df} with 0 and convert to integers\;
\textbf{Sort} \texttt{df} by \texttt{sample\_id} in ascending order\;

\Return{\texttt{df}}
\end{algorithm}

\begin{table*}[htpb]
\centering
\caption{ Summary of GUIDE-MLRan and MLRan Dataset Compliance with the Guidelines. It outlines the fifteen criteria of the GUIDE-MLRan framework for constructing high-quality behavioural ransomware datasets. Columns indicate the guideline category, specific criterion (C1–C15), a brief description of the requirement, and a summary of how the MLRan dataset fulfils each criterion.}
\label{tab:compliance_main}
\resizebox{\textwidth}{!}{%
\begin{tabular}{llll}
\hline
\multicolumn{1}{c}{\textbf{Category}} &
  \multicolumn{1}{c}{\textbf{Criteria (C)}} &
  \textbf{Description} &
  \multicolumn{1}{c}{\textbf{MLRan Compliance Summary}} \\ \hline
\multirow{3}{*}{\textbf{\begin{tabular}[c]{@{}l@{}}Sample Diversity \\ and \\ Representativeness\end{tabular}}} &
  \begin{tabular}[c]{@{}l@{}}C1. Diverse Ransomware\\ Samples\end{tabular} &
  \begin{tabular}[c]{@{}l@{}}Include varied ransomware types and \\ families (e.g., Locker, Crypto, RaaS, Modern).\end{tabular} &
  \begin{tabular}[c]{@{}l@{}}Includes 2,330 samples across 64 families, covering\\  all major ransomware types (Crypto, RaaS, Locker, \\ Modern), reflecting both legacy and emerging variants.\end{tabular} \\ \cline{2-4} 
 &
  \begin{tabular}[c]{@{}l@{}}C2. Diverse Goodware \\ Samples\end{tabular} &
  \begin{tabular}[c]{@{}l@{}}Include diverse benign software (e.g., common \\ applications, browsers, word processors, utilities).\end{tabular} &
  \begin{tabular}[c]{@{}l@{}}Curated 2,550 goodware samples across 11 real-world \\ categories, including productivity, developer tools, \\ internet utilities, games, and security software.\end{tabular} \\ \cline{2-4} 
 &
  \begin{tabular}[c]{@{}l@{}}C3. Balanced Class \\ Distribution\end{tabular} &
  \begin{tabular}[c]{@{}l@{}}Maintain a balanced ratio of ransomware and\\  goodware samples.\end{tabular} &
  \begin{tabular}[c]{@{}l@{}}Achieves inter-class balance (52.25\% goodware, 47.75\%\\ ransomware); intra-class diversity captured via broad \\ ransomware family representation.\end{tabular} \\ \hline
\multirow{3}{*}{\textbf{\begin{tabular}[c]{@{}l@{}}Sample Quality \\ and Accuracy\end{tabular}}} &
  C4. Accurate Labelling &
  \begin{tabular}[c]{@{}l@{}}Ensure accurate annotation of ransomware and \\ goodware with detailed metadata.\end{tabular} &
  Ground truth family labels and naming using AVClass. \\ \cline{2-4} 
 &
  C5. Sample size &
  \begin{tabular}[c]{@{}l@{}}Collect enough samples to capture common \\ ransomware behavioural patterns.\end{tabular} &
  \begin{tabular}[c]{@{}l@{}}4880 samples sourced from four high-quality repositories (EldeRan, \\ MOTIF, MarauderMap, and curated threat intel); \\ ensures comprehensive, high-fidelity behavioural coverage.\end{tabular} \\ \cline{2-4} 
 &
  C6. Covered Time Period &
  \begin{tabular}[c]{@{}l@{}}Include both recent and outdated ransomware \\ for real-world applicability.\end{tabular} &
  \begin{tabular}[c]{@{}l@{}}Samples span from 2006–2024 using VirusTotal first \\ submission timestamps.\end{tabular} \\ \hline
\multirow{2}{*}{\textbf{\begin{tabular}[c]{@{}l@{}}Sandbox and \\ Testbed \\ Requirements\end{tabular}}} &
  C7. Sandbox Hardening &
  \begin{tabular}[c]{@{}l@{}}Simulate real-world user behaviour to bypass \\ anti-sandbox techniques.\end{tabular} &
  \begin{tabular}[c]{@{}l@{}}Cuckoo sandbox hardened using disguise and human \\ activity simulation modules, extended timeouts, and \\ realistic system configurations for evasion countermeasures.\end{tabular} \\ \cline{2-4} 
 &
  \begin{tabular}[c]{@{}l@{}}C8. Comprehensive \\ Behavioural \\ Analysis\end{tabular} &
  \begin{tabular}[c]{@{}l@{}}Capture dynamic and static features for a \\ comprehensive behavioural profile.\end{tabular} &
  \begin{tabular}[c]{@{}l@{}}Extracts nine dynamic behavioural feature groups \\ (API, REG, FILE, DIR, STR, NET, SYS, DROP, SIG) \\ from Cuckoo reports for robust malware profiling.\end{tabular} \\ \hline
\multirow{3}{*}{\textbf{\begin{tabular}[c]{@{}l@{}}Representative \\ Feature Extraction \\ and Modelling\end{tabular}}} &
  \begin{tabular}[c]{@{}l@{}}C9. Relevant Feature \\ Extraction\end{tabular} &
  \begin{tabular}[c]{@{}l@{}}Extract features such as API calls, file operations \\ and network behaviours relevant to detection.\end{tabular} &
  \begin{tabular}[c]{@{}l@{}}Publicly available scripts extract critical features from \\ structured JSON reports; features are relevant to file, \\ system, and network activity.\end{tabular} \\ \cline{2-4} 
 &
  C10. Data Preprocessing &
  \begin{tabular}[c]{@{}l@{}}Clean, normalise, and standardise data to \\ ensure quality and consistency.\end{tabular} &
  \begin{tabular}[c]{@{}l@{}}Preprocessing includes duplicate removal, standardisation,\\  binary encoding, and metadata alignment for consistent ML input.\end{tabular} \\ \cline{2-4} 
 &
  \begin{tabular}[c]{@{}l@{}}C11. Model Training \\ and Evaluation\end{tabular} &
  Ensure robust model training and evaluation. &
  \begin{tabular}[c]{@{}l@{}}Models trained using rigorous protocol to avoid test/temporal/\\ selective snooping bias; evaluation incorporates several \\ performance evaluation metrics.\end{tabular} \\ \hline
\multirow{4}{*}{\textbf{\begin{tabular}[c]{@{}l@{}}Documentation, \\ Reproducibility, \\ and Data Extension\end{tabular}}} &
  \begin{tabular}[c]{@{}l@{}}C12. Availability of \\ Contextual and \\ Metadata Information\end{tabular} &
  \begin{tabular}[c]{@{}l@{}}Provide detailed metadata about samples, \\ datasets, sandboxes and methodologies.\end{tabular} &
  \begin{tabular}[c]{@{}l@{}}Comprehensive metadata provided (hashes, types, timestamps, \\ labels, source); sandbox configuration and feature definitions\\  clearly documented.\end{tabular} \\ \cline{2-4} 
 &
  \begin{tabular}[c]{@{}l@{}}C13. Public Availability \\ of Data and Code\end{tabular} &
  \begin{tabular}[c]{@{}l@{}}Share datasets and codes with documentation to \\ encourage reproducibility and benchmarking.\end{tabular} &
  \begin{tabular}[c]{@{}l@{}}All Cuckoo parsers, automation scripts, and dataset metadata \\ made available on GitHub with documentation and version \\ control.\end{tabular} \\ \cline{2-4} 
 &
  \begin{tabular}[c]{@{}l@{}}C14. Ethical and Legal \\ Considerations\end{tabular} &
  \begin{tabular}[c]{@{}l@{}}Ensure anonymisation, privacy, and compliance \\ with regulations.\end{tabular} &
  \begin{tabular}[c]{@{}l@{}}Anonymisation ensured; only metadata and derived features shared; \\ binaries handled securely and access-controlled to prevent misuse.\end{tabular} \\ \cline{2-4} 
 &
  C15. Continuous Updates &
  \begin{tabular}[c]{@{}l@{}}Regularly update datasets to include emerging \\ ransomware threats.\end{tabular} &
  \begin{tabular}[c]{@{}l@{}}Dataset incorporates ransomware samples up to 2024; framework \\ enables longitudinal extension and collaborative community curation.\end{tabular} \\ \hline
\end{tabular}
}
\end{table*}

\section{Experimental design} \label{sec:experimental_design}

This section presents our approach to conducting the experiments demonstrating the effectiveness of the MLRan dataset. It illustrates the research questions and the process adopted to answer them, the selected feature selection techniques and ML models, and the evaluation metrics we used to evaluate ransomware detection performance.

\subsection{Research questions}
In this study, we aim to answer the following research questions:
\begin{itemize}
    \item RQ1: How effectively do traditional ML classifiers distinguish ransomware from goodware using the MLRan dataset?
    \item RQ2: How do feature selection techniques impact the performance of ransomware detection models on the MLRan dataset?
    \item RQ3: Which features in the MLRan dataset are most predictive for ransomware detection?
    \item RQ4: How can analysing misclassified decisions reveal model weaknesses, enhance interpretability, and improve the robustness of ransomware detection models?
\end{itemize}

\subsection{Feature selection techniques}

We employed feature selection (FS) techniques to ensure unique feature names, eliminate duplicate entries, and remove constant features lacking variance.
Feature selection identifies and selects the most relevant features from a dataset, reducing dimensionality while preserving the model's decision-making quality \citep{zebari2020comprehensive}. Eliminating redundant or irrelevant variables simplifies models, improves computational efficiency, enhances model interpretability, reduces overfitting, and ultimately increases the accuracy and generalisability of ML algorithms \citep{theng2024feature}.

Feature selection methods fall into four categories: filter, wrapper, embedded methods, and genetic algorithms \citep{kamolov2021feature}. Filter methods using measures like information gain or $\chi^2$-statistics to evaluate features independently are the most straightforward and efficient. Wrapper methods evaluate feature subsets by training a model, often giving better results but requiring more computation, such as Recursive Feature Elimination (RFE). Embedded methods combine feature selection with the learning process, usually through regularisation. Genetic algorithms select features by evolving subsets based on their performance, offering a flexible, optimisation-based approach. While wrapper and embedded methods may work better with specific models, filter methods are more consistent across different ones.

This study employs three widely used techniques: Mutual Information, Chi-Square, and RFE.  

\subsubsection{Mutual Information (MI)}
Mutual Information quantifies the dependency between a feature \(X\) and the target variable \(Y\), measuring the reduction in uncertainty about \(Y\) given \(X\):  

\begin{equation}
I(X; Y) = \sum_{x \in X} \sum_{y \in Y} P(x, y) \log \frac{P(x, y)}{P(x) P(y)}
\end{equation}

A higher \(I(X; Y)\) value indicates stronger feature relevance. MI is non-parametric and captures both linear and nonlinear relationships.  

\subsubsection{Chi-Square (\(\chi^2\)) Test}  
The \(\chi^2\) test evaluates the statistical dependence between a categorical feature and the target variable, computed as:  

\begin{equation}
\chi^2 = \sum_{i=1}^{n} \frac{(O_i - E_i)^2}{E_i}
\end{equation}

where \( O_i \) and \( E_i \) denote the observed and expected frequencies, respectively. A high \(\chi^2\) score suggests strong dependence, making the feature valuable for classification.  

\subsubsection{Recursive Feature Elimination (RFE)}  
RFE iteratively removes the least important features based on model coefficients (e.g., \( \beta \) in linear models) or feature importance scores from tree-based models. The process refines feature selection by retaining those with the highest predictive contribution.

\subsection{Machine learning models}


Machine learning models vary in complexity and interpretability, offering different performance and computational efficiency trade-offs. This study employs Light Gradient Boosting Machine (LightGBM) , Decision Trees, Logistic Regression, and Random Forest, each with distinct advantages in ransomware detection.

\subsubsection{Light gradient boosting machine (LightGBM)}  
LightGBM is a gradient boosting framework that constructs decision trees using a leaf-wise growth strategy, which selects the leaf with the maximum loss reduction to split. This approach often leads to faster convergence and better accuracy than level-wise methods. LightGBM is optimised for high performance through histogram-based binning, gradient-based one-side sampling (GOSS), and exclusive feature bundling (EFB). The \textit{objective function} follows the regularised gradient boosting formulation:

\begin{equation}
\mathcal{L}(\Theta) = \sum_{i=1}^{n} l(y_i, \hat{y}_i) + \lambda \sum_{k=1}^{K} \|f_k\|^2
\end{equation}

where \( l(y_i, \hat{y}_i) \) is the differentiable loss function, and the regularisation term \( \|f_k\|^2 \) controls tree complexity. LightGBM is particularly effective for large-scale and high-dimensional learning tasks such as ransomware classification.

\subsubsection{Decision trees}  
A Decision Tree splits data using conditions at nodes, forming a tree structure for decision-making. It uses measures like \textit{Gini impurity}:

\begin{equation}
G = 1 - \sum_{i=1}^{C} p_i^2
\end{equation}

where \( p_i \) is the probability of a class. Decision Trees are interpretable but prone to overfitting.

\subsubsection{Logistic regression}  
Logistic Regression models the probability of a binary outcome using the \textit{sigmoid function}:

\begin{equation}
P(Y=1 | X) = \frac{1}{1 + e^{-(\beta_0 + \sum_{i=1}^{n} \beta_i x_i)}}
\end{equation}

where \( \beta_0 \) is the intercept and \( \beta_i \) are feature coefficients. It is computationally efficient but limited to linearly separable data.

\subsubsection{Random forest}  
Random Forest is a bagging ensemble of Decision Trees, reducing overfitting by averaging multiple tree predictions. The final classification is determined via \textit{majority voting}:

\begin{equation}
\hat{y} = \text{mode}(y_1, y_2, ..., y_T)
\end{equation}

where \( y_T \) is the prediction from the \( T \)-th tree. For regression, predictions are averaged:

\begin{equation}
\hat{y} = \frac{1}{T} \sum_{t=1}^{T} y_t
\end{equation}

It handles high-dimensional data effectively but is computationally expensive.

\subsubsection{Extra trees (Extremely randomised trees)}  
Extra Trees is an ensemble method that builds multiple unpruned decision trees using randomly selected features and split thresholds. Unlike Random Forests, it increases randomness to reduce variance and enhance generalisation. Predictions are aggregated as:

\begin{equation}
\hat{y} = \frac{1}{T} \sum_{t=1}^{T} f_t(x)
\end{equation}

where \( f_t(x) \) is the prediction from the \( t^{th} \) tree. Extra Trees is efficient and effective for high-dimensional tasks such as ransomware detection.


\subsection{Evaluation metrics}


To assess model performance in ransomware detection, we employed the following evaluation metrics:

\subsubsection{Accuracy}  
Accuracy measures the proportion of correctly classified instances:

\begin{equation}
\text{Accuracy} = \frac{TP + TN}{TP + TN + FP + FN}
\end{equation}

where \( TP \) and \( TN \) represent true positives and true negatives, respectively, while \( FP \) and \( FN \) denote false positives and false negatives. Accuracy is effective for balanced datasets but may be misleading in imbalanced scenarios.

\subsubsection{Balanced accuracy}  
Balanced Accuracy adjusts for class imbalances by averaging the recall of each class:

\begin{equation}
\text{Balanced Accuracy} = \frac{1}{2} \left( \frac{TP}{TP + FN} + \frac{TN}{TN + FP} \right)
\end{equation}

This metric ensures a fair evaluation when class distributions are skewed.

\subsubsection{Precision}  
Precision quantifies the proportion of correctly identified positive instances out of all predicted positives:

\begin{equation}
\text{Precision} = \frac{TP}{TP + FP}
\end{equation}

A high precision indicates fewer false positives, which is critical in security applications where misclassifying benign software as ransomware has significant consequences.

\subsubsection{Recall}  
Recall, or sensitivity, measures the model’s ability to detect positive instances:

\begin{equation}
\text{Recall} = \frac{TP}{TP + FN}
\end{equation}

A high recall ensures fewer false negatives, making it essential for accurately identifying ransomware threats.

\subsubsection{F1-score}  
The F1-Score balances precision and recall, providing a single performance metric:

\begin{equation}
F1 = 2 \times \frac{\text{Precision} \times \text{Recall}}{\text{Precision} + \text{Recall}}
\end{equation}

It is particularly useful when false positives and false negatives need equal consideration.

\subsection{Ransomware detection pipeline}

\begin{figure*}[htpb]
    \centering
    \includegraphics[width=1\textwidth]{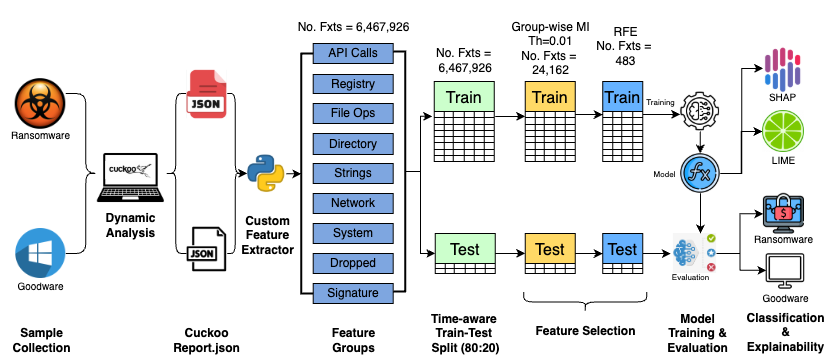}
    \caption{Ransomware detection pipeline showing dynamic analysis, custom feature extraction, multi-stage feature selection, model training, and explainability using SHAP and LIME.}%
    \label{fig:model_design}
\end{figure*}

To answer the research questions, we performed the activities in the ransomware detection pipeline presented in Figure \ref{fig:model_design}. First, we conducted the dynamic analysis of the collected ransomware and goodware samples using Cuckoo Sandbox, which generates JSON reports capturing runtime behaviour. Then, we used a custom  Python-based extractor to parse these reports to generate over 6.4 million binary features, grouped into nine behavioural categories, including API calls, registry activity, file operations, and network behaviour. 

A \textbf{time-aware 80:20 train-test split} ensures temporal consistency and realistic evaluation. Specifically, samples were separated by their types and chronologically ordered by their first submission date to VirusTotal, and the earliest 80\% are used for training while the most recent 20\% form the test set. This strategy prevents temporal leakage and mimics real-world deployment, where models are trained on past data and evaluated on future, unseen samples. Such temporal separation is crucial for assessing model generalisation, especially in the context of evolving ransomware behaviours and concept drift.

Then, we performed feature selection in two stages. First, we conducted group-wise mutual information filtering (threshold = 0.01), which reduced the feature space from 6.4 million to 24,162. Then we performed RFE, which selects the most informative 483 features. We then trained a classification model on these features and evaluated it on the test set.

To enhance model transparency, we applied SHAP and LIME  to explain individual predictions and highlight influential features, providing interpretable insights into the behavioural traits that distinguish ransomware from goodware.

\section{Results} 
\label{sec:results}
This section presents the results we obtained to answer the research questions:  performance of traditional ML classifiers (baselines) (RQ1), the impact of feature selection on the performance of ransomware detection models (RQ2), feature importance (RQ3), analysis of model decisions, and explanations of misclassified instances at both global and local levels (RQ4).


\subsection{Baseline performance (RQ1)}


We compared the performance of the baseline ML models for distinguishing ransomware from goodware using  three feature selection techniques.Table~\ref{tab:bin_fs} presents our comparative analysis. We consider three FS methods as part of stage 1 (MI with thresholds of 0.01 and 0.001, and Chi-Square) highlighting trade-offs between model performance and computational cost. While logistic regression achieves marginally higher accuracy and F1 score with MI at a 0.001 threshold, MI at 0.01 is selected as the preferred method due to its ability to retain significantly fewer features while delivering comparable performance (97.54\% accuracy and F1 score). Notably, it also achieves the lowest computation time (4.75 seconds), making it the most efficient choice. This threshold reduces the feature space from 6,467,926 to 24,162, enhancing model interpretability and reducing complexity without compromising predictive power. Overall, MI-based feature selection consistently outperforms Chi-Square, and logistic regression emerges as the most effective and computationally efficient model. In contrast, LightGBM performs the least favourably across all FS methods, particularly with MI (0.001), where it exhibits both lower accuracy and significantly higher computation time, indicating poor scalability in this setting.

\begin{tcolorbox}[colback=lightash, colframe=black!50!black, title=Result 1: Baseline performance] 
Logistic regression with MI (0.01) offers the best trade-off between performance, efficiency, and feature reduction, while LightGBM performs worst due to lower accuracy and high computation cost.
\end{tcolorbox}

\begin{table*}[htpb]
\caption{Performance comparison of machine learning models for ransomware and goodware detection using three feature selection (FS) techniques: Mutual Information (MI) with thresholds of 0.01 and 0.001, and Chi-Square. The table presents accuracy (Acc), balanced accuracy (Bal. Acc), precision (Pre), recall (Re), F1 score (F1), and computation time in seconds (Time) for each model: XGBoost, Decision Trees, Logistic Regression, and Random Forest. Results show that logistic regression with MI (0.01) outperforms other FS methods in terms of performance and computational efficiency.}
\label{tab:bin_fs}
\begin{tabular}{llcccccc}
\hline
\textbf{FS} & \textbf{Model}               & \textbf{Acc}   & \textbf{Bal. Acc} & \textbf{Pre}   & \textbf{Re}    & \multicolumn{1}{l}{\textbf{F1}} & \textbf{Time} \\ \hline
\multirow{4}{*}{\textbf{MI (0.01)}}  & LightGBM             & 94.05 & 94.07 & 94.06 & 94.05 & 94.05 & 7876.11  \\

                                     & Decision Trees      & 97.54 & 97.59 & 97.57 & 97.54 & 97.54 & 10.76  \\
                                 
            & Logistic Regression & 97.54 & 97.59    & 97.57 & 97.54 & 97.54                 & 4.75 \\
                                     & Random Forest       & 97.03 & 97.10 & 97.09 & 97.03 & 97.03 & 13.41   \\ 
                                            & Extra Trees      & 97.33 & 97.38 & 97.37 & 97.33 & 97.33 & 4.98   \\                              \hline
                           
\multirow{4}{*}{\textbf{MI (0.001)}} & LightGBM             & 94.56 & 94.56 & 94.56 & 94.56 & 94.56 & 24797.11 \\

                                     & Decision Trees      & 96.62 & 96.67 & 96.65 & 96.62 & 96.62 & 50.89  \\

                                     & Logistic Regression & 98.05 & 98.11 & 98.05 & 98.05 & 98.05 & 19.97   \\
                                     
                                     & Random Forest       & 96.92 & 97.00 & 96.99 & 96.92 & 96.92 & 37.40   \\                              
                                         & Extra Trees      & 96.92 & 96.99 & 96.98 & 96.92 & 96.92 & 20.68   \\ \hline
\multirow{4}{*}{\textbf{Chi-Square}} & LightGBM             & 93.84 & 93.85 & 93.85 & 93.85 & 98.41 & 23017.73 \\
                                     & Decision Trees      & 96.82 & 96.86 & 96.85 & 96.82 & 96.82 & 115.36  \\
                                     & Logistic Regression & 98.05 & 98.11 & 98.09 & 98.05 & 98.05 & 120.10  \\
                                     & Random Forest       & 93.54 & 93.79  & 94.13 & 93.54 & 93.53 & 968.32  \\ 
                                         & Extra Trees       & 93.54 & 93.80 & 94.17 & 93.54 & 93.53 & 833.34   \\ \hline
\end{tabular}%
\end{table*}

\subsection{Impact of feature selection (RQ2)}

\begin{figure}[htpb]
\centering 
  \includegraphics[width=0.99\columnwidth]{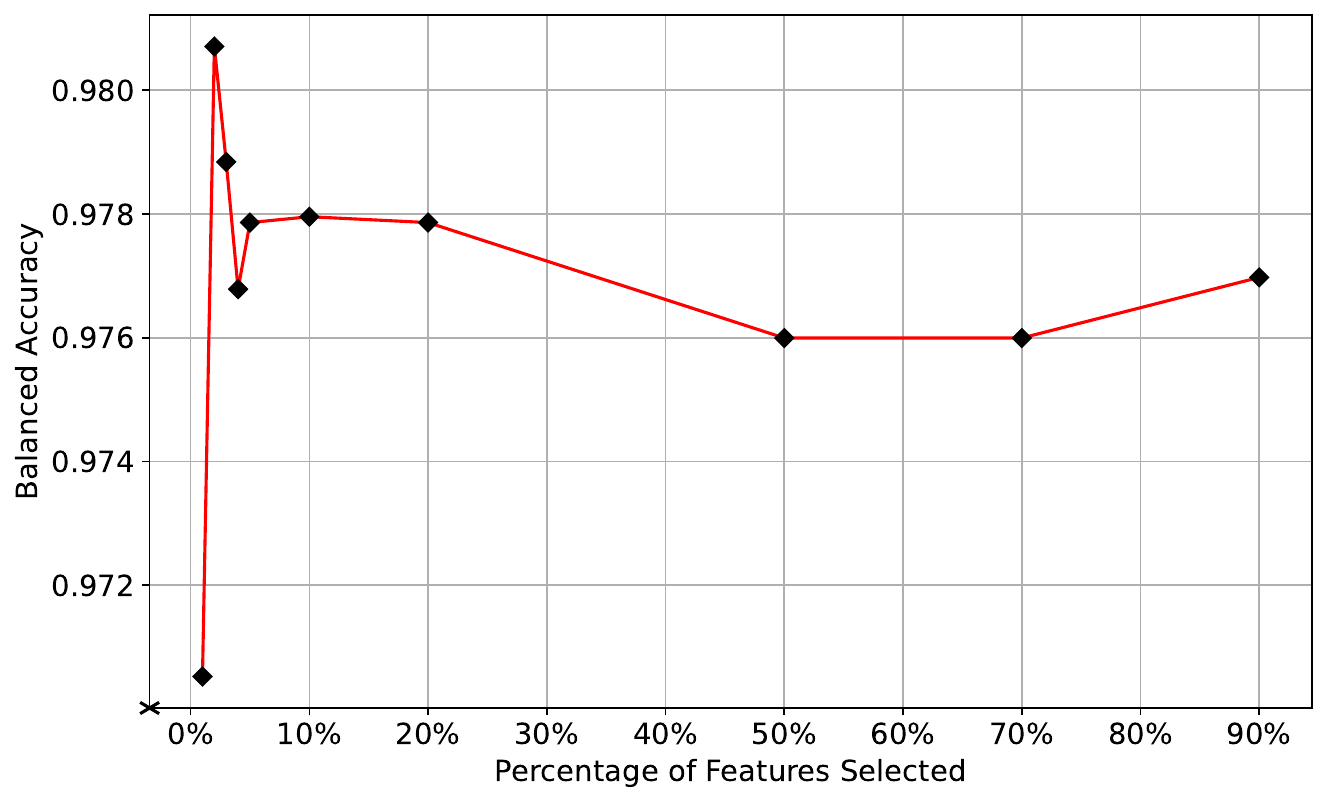}%
\caption{This plot shows the balanced accuracy of a logistic regression model as a function of the percentage of features selected using Recursive Feature Elimination. The initial dataset contains 24,162 features, and RFE is used to select subsets of features corresponding to different percentages iteratively. The first 5 points on the x-axis are  1\%, 2\%, 3\%, 4\%, and 5\%. The highest balanced accuracy (98.07\%) is achieved when only 2\% of the features (483 features) are selected, demonstrating that careful feature selection can improve model performance while reducing computational complexity. }
\label{fig:rfe_balance}
\end{figure}

In Stage 1 of our feature selection pipeline, MI (0.01) yielded an initial feature set of 24,162. In Stage 2, this set was reduced to 483 features using RFE as described in Algorithm \ref{algo:rfe}, with logistic regression (the best-performing model from Stage 1) as the base estimator. The algorithm iteratively evaluates performance across varying feature counts: it initialises a logistic regression model (Lines 1–2), computes feature counts from selection percentages (Lines 5–6), applies RFE (Lines 11–14), trains and tests the model (Lines 15–18), and records balanced accuracy (Lines 19–21). Results are stored and returned in a dictionary (Lines 22–25). Figure \ref{fig:rfe_balance} shows that the highest balanced accuracy (98.07\%) is achieved using just 2\% of the features (483), highlighting the effectiveness of the proposed feature selection technique.

Table \ref{tab:rfe_fs} presents the performance of the models on the selected 483 features across three ransomware detection tasks: binary, multi-class (type), and family classification. Logistic Regression delivers strong, consistent results with high accuracy and minimal computation time, particularly in the binary and family classifications. Extra trees model performs best in ransomware type classification. LightGBM shows lower balanced accuracy on the binary and family classification tasks.




\begin{algorithm}[htbp]
\SetAlgoLined
\KwIn{$X_{\text{train}}$: Training features, $X_{\text{test}}$: Testing features, \\
$y_{\text{train}}$: Training labels, $y_{\text{test}}$: Testing labels}
\KwOut{A dictionary mapping percentages to balanced accuracy scores}

\textbf{Step 1: Initialise logistic regression model} \\
Set $model \gets $ \text{LogisticRegression(max\_iter=10000,} \\
\text{solver='saga', random\_state=42)}\\

\textbf{Step 2: Define feature selection percentages} \\
Set $percentages \gets [1, 2, 3, 4, 5, 10, 20, 50, 70, 90]$ \\
Set $feature\_counts \gets [\lfloor (p / 100) \cdot |X_{\text{train}}| \rfloor \text{ for } p \in percentages]$ \\

\textbf{Step 3: Initialise balanced accuracies dictionary} \\
Set $balanced\_accuracies \gets \{\}$ \\

\textbf{Step 4: Recursive Feature Elimination (RFE)} \\
\ForEach{$num\_features \in feature\_counts$}{
    \textbf{Perform RFE for $num\_features$ features:} \\
    Set $rfe \gets $ \text{RFE(estimator=model,}\\ $\text{n\_features\_to\_select=num\_features, step=0.1)}$ \\
    Fit RFE: $rfe.fit(X_{\text{train}}, y_{\text{train}})$ \\

    \textbf{Transform test set:} \\
    Set $X_{\text{test\_transformed}} \gets rfe.transform(X_{\text{test}})$ \\

    \textbf{Fit logistic regression on selected features:} \\
    Fit $model$ using $rfe.transform(X_{\text{train}})$ and $y_{\text{train}}$ \\

    \textbf{Calculate balanced accuracy:} \\
    Set $y_{\text{pred}} \gets model.predict(X_{\text{test\_transformed}})$ \\
    Set $balanced\_acc \gets \text{balanced\_accuracy\_score}(y_{\text{test}}, y_{\text{pred}})$ \\

    Add to dictionary: $balanced\_accuracies[num\_features] \gets balanced\_acc$ \\
}

\textbf{Step 5: Return balanced accuracies dictionary} \\
\Return $balanced\_accuracies$
\caption{Perform Recursive Feature Elimination and evaluate logistic regression performance}
\label{algo:rfe}
\end{algorithm}

\begin{table*}[htpb]
\centering
\caption{Performance comparison of machine learning models for ransomware classifications across the three labels using the features from RFE. The table presents accuracy (Acc), balanced accuracy (Bal. Acc), precision (Pre), recall (Re), F1 score (F1), and computation time (Time) for each model: XGBoost, Decision Trees, Logistic Regression, and Random Forest. }
\label{tab:rfe_fs}
\begin{tabular}{llcccccc}
\hline
\textbf{FS} & \textbf{Model} & \textbf{Acc} & \textbf{Bal. Acc} & \textbf{Pre} & \textbf{Re} & \multicolumn{1}{l}{\textbf{F1}} & \textbf{Time} \\ \hline
\multirow{4}{*}{\textbf{Binary}}   & LightGBM             & 94.05 & 94.02 & 94.05 & 94.05 & 94.05 & 1206.58  \\
                                   & Decision Trees      & 93.54 & 93.43 & 93.61  & 93.54 & 93.53 & 12.13  \\
                                   & Logistic Regression & 98.15 & 98.18 & 98.16 & 98.15 & 98.15 & 0.86  \\
                                   & Random Forest       & 96.62 & 96.63 & 96.62 & 96.62 & 96.62 & 2.21  \\ 
                                    & Extra Trees      & 97.13 & 97.18 & 97.16 & 97.13 & 97.13 & 1.60  \\                                     \hline
\multirow{4}{*}{\textbf{Types}}    & LightGBM             & 81.95 & 73.43  & 86.61 & 82.34 & 81.47 & 5447.46  \\
                                   & Decision Trees      & 83.79 & 74.43 & 84.43 & 83.79 & 83.78  & 12.17    \\
                                   & Logistic Regression & 86.46 & 79.86 & 87.70 & 86.46 & 86.53 & 1.31    \\
                                   & Random Forest       & 86.67 & 79.34  & 87.43 & 86.67 & 86.65 & 2.67    \\ 
                                    & Extra Trees      & 87.69 & 80.71 & 88.32 & 87.69 & 87.77 & 1.61  \\                                  \hline
\multirow{4}{*}{\textbf{Families}} & LightGBM             & 74.15 & 30.42  & 69.06  & 74.15 & 68.07 & 7132.76\\
                                   & Decision Trees      & 75.38 & 41.66    & 80.84 & 75.38 & 76.72 & 1.79    \\
                                   & Logistic Regression & 81.85 & 47.07 & 83.70 & 81.85 & 81.35 & 2.24    \\
                                   & Random Forest       & 78.77 & 45.55 & 80.82 & 78.77 & 77.84 & 2.95   \\ 
                                     & Extra Trees      & 81.13 & 45.23 & 83.35 & 81.13 & 81.31 & 1.54  \\                                    \hline
\end{tabular}
\end{table*}

\begin{table}[htpb]
\caption{Distribution of number of features selected across the feature groups in the MLRan dataset: application programming interface (API), registry keys (REG), file operations (FILE), directory operations (DIR), strings (STR), network activities (NET), system operations (SYS), drop operations (DROP), and signatures (SIG). Feature selection is carried out in two stages: Stage 1 applies mutual information (MI) with a 0.01 threshold to the original features, and Stage 2 uses recursive feature elimination (RFE) on the features selected in Stage 1. N represents the number of features belonging to that group and \% represents the percentage of each feature group relative to the total features selected by each technique.}
\label{tab:fs_group}
\resizebox{\columnwidth}{!}{%
\begin{tabular}{lcc|cc|cc}
\hline
\multicolumn{1}{c}{\multirow{2}{*}{\textbf{\begin{tabular}[c]{@{}c@{}}Feature\\ Groups\end{tabular}}}} &
  \multicolumn{2}{c|}{\textbf{Original}} &
  \multicolumn{2}{c|}{\textbf{After MI(0.01) FS}} &
  \multicolumn{2}{c}{\textbf{After RFE FS}} \\ \cline{2-7} 
\multicolumn{1}{c}{} &
  \textbf{N} &
  \textbf{\%} &
  \textbf{N} &
  \textbf{\%} &
  \textbf{N} &
  \textbf{\%} \\ \hline
\textbf{API}  & 313     & 0.0048  & 137   & 0.5670  & 63  & 13.0434 \\
\textbf{REG}  & 525505  & 8.1248  & 5369  & 22.2208 & 90  & 18.6335 \\
\textbf{FILE} & 2078287 & 32.1322 & 14044 & 58.1243 & 16  & 3.3126  \\
\textbf{DIR}  & 158123  & 2.4447  & 820   & 3.3938  & 10  & 2.0704  \\
\textbf{STR}  & 3632119 & 56.1559 & 3347  & 13.8523 & 223 & 46.1697 \\
\textbf{NET}  & 4814    & 0.0744  & 3     & 0.0124  & 0   & 0.0000  \\
\textbf{SYS}  & 16911   & 0.2615  & 321   & 1.3285  & 38  & 7.8674  \\
\textbf{DROP} & 51651   & 0.7986  & 75    & 0.3104  & 23  & 4.7619  \\
\textbf{SIG}  & 203     & 0.0031  & 46    & 0.1904  & 20  & 4.1407 \\ \hline
Total         & 6467926 &         & 24162 &         & 483 &         \\ \hline
\end{tabular}%
}
\end{table}

Table \ref{tab:fs_group} shows the number of features selected in each feature group at each stage of the two stage feature selection technique. In the \textbf{API} feature group, the proportion of selected features increases significantly from 0.0048\% in the original dataset to 0.5670\% after MI(0.01) and further to 13.0434\% after RFE, highlighting the increasing importance of API features. Similarly, for \textbf{REG} (Registry Keys), the proportion rises from 8.1248\% to 22.2208\% after MI(0.01), then drops slightly to 18.6335\% after RFE, reflecting the selection of more relevant features during the MI(0.01) stage, followed by refinement with RFE.

In the \textbf{FILE} feature group, MI(0.01) increases the selected proportion from 32.1322\% to 58.1243\%, but RFE significantly reduces it to 3.3126\%, indicating that RFE retained only the most relevant file-related features. The \textbf{DIR} (Directory Operations) group shows a modest increase from 2.4447\% to 3.3938\% after MI(0.01), followed by a small reduction to 2.0704\% after RFE, suggesting less importance placed on directory operations.

The \textbf{STR} (Strings) feature group shows a dramatic shift, with the percentage dropping from 56.1559\% to 13.8523\% after MI(0.01) and then increasing to 46.1697\% after RFE, indicating that RFE preserved a significant number of string-based features. In contrast, \textbf{NET} (Network Activities) features were reduced from 0.0744\% to 0.0124\% after MI(0.01) and completely eliminated (0\%) after RFE, suggesting their limited relevance to the model.

For \textbf{SYS} (System Operations), the proportion increases from 0.2615\% to 1.3285\% after MI(0.01) and increases to 7.8674\% after RFE. The \textbf{DROP} (Drop Operations) group shows a decrease from 0.7986\% to 0.3104\% after MI(0.01) and a slight increase to 4.7619\% after RFE, highlighting RFE’s role in retaining important drop operations.

Finally, \textbf{SIG} (Signatures) features increase from 0.0031\% to 0.1904\% after MI(0.01) and then to 4.1407\% after RFE, indicating that only the most critical signature features were retained.

\begin{tcolorbox}[colback=lightash, colframe=black!50!black, title=Result 2: Impact of feature selection] 
The proposed two-stage feature selection method reduced over 6.4 million features to 483, significantly improving model efficiency and interoperability.

In addition, Table~\ref{tab:fs_group} shows how feature selection reshapes the distribution of behavioural feature groups. While \texttt{STR} and \texttt{FILE} features initially dominate the dataset, comprising over 88\% of all features, recursive feature elimination reduces the prominence of \texttt{STR} (46.58\%) and \texttt{FILE} to just 3.31\%. However, the relative importance of \texttt{REG} and \texttt{API} features increased, reflecting their relevance to ransomware detection. In contrast, \texttt{NET} features are entirely removed, indicating limited discriminative value. Overall, the process highlights string, registry, API and system operation behaviours as the most informative.
\end{tcolorbox}


\subsection{Feature importance (RQ3)}


Global feature importance refers to the overall contribution of each feature to a model's decision-making process across the entire dataset. It quantifies how much each feature influences the model's predictions, providing insights into which features are most critical for the model’s behaviour. In this study, we utilised SHAP (SHapley Additive exPlanations). SHAP is a powerful and widely used method for model interpretability that provides a unified framework for explaining predictions of machine learning models. It is based on Shapley values, a concept borrowed from cooperative game theory, which assigns each feature an importance value based on its contribution to the model's output.

\begin{figure*}[htpb]
\centering 
  \includegraphics[width=1\textwidth]{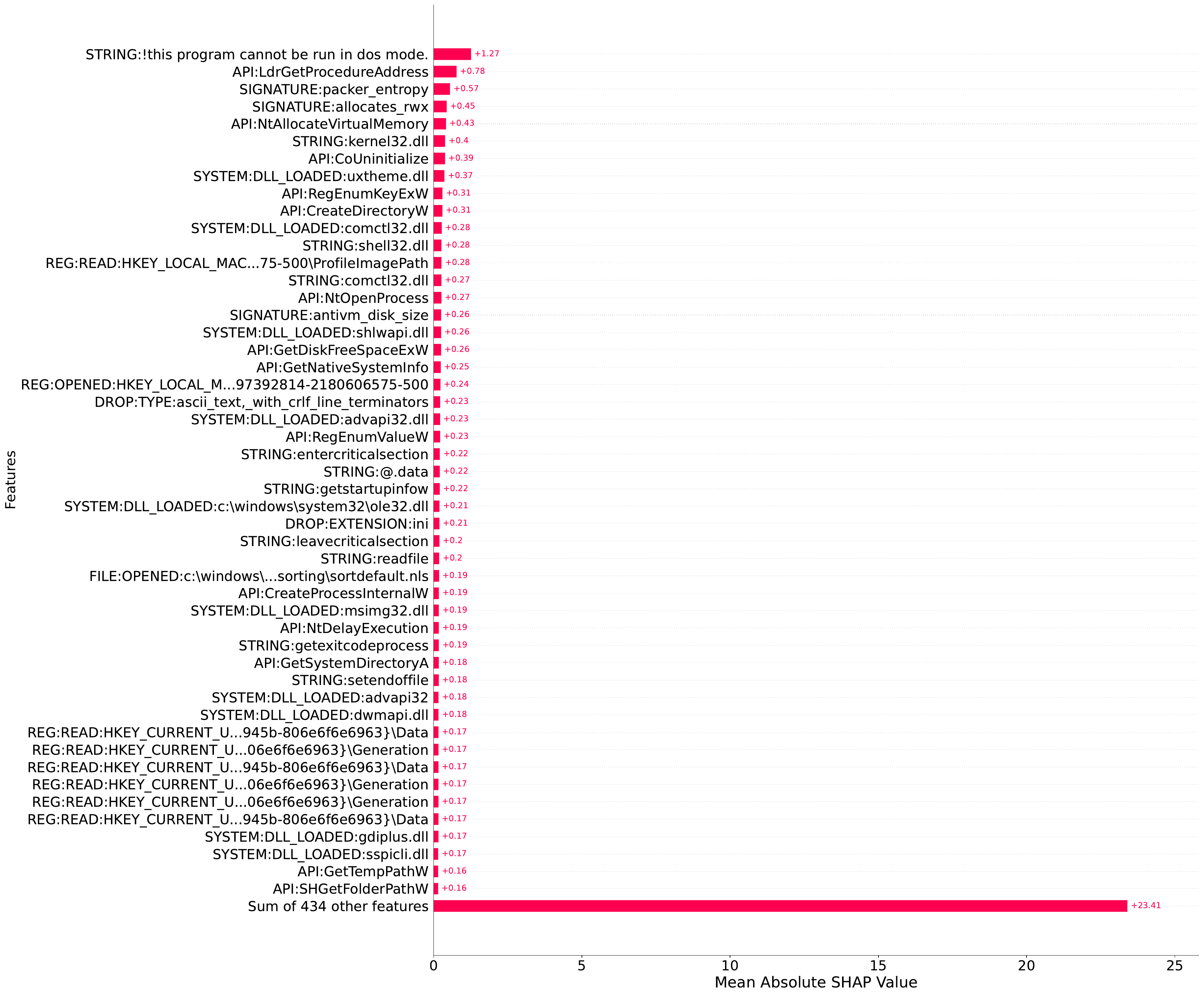}%
\caption{SHAP decision plot showing the top 50 features ranked by their impact on model output. }
\label{fig:shap_bar}
\end{figure*}

\begin{figure*}[htpb]
\centering 
  \includegraphics[scale=0.3]{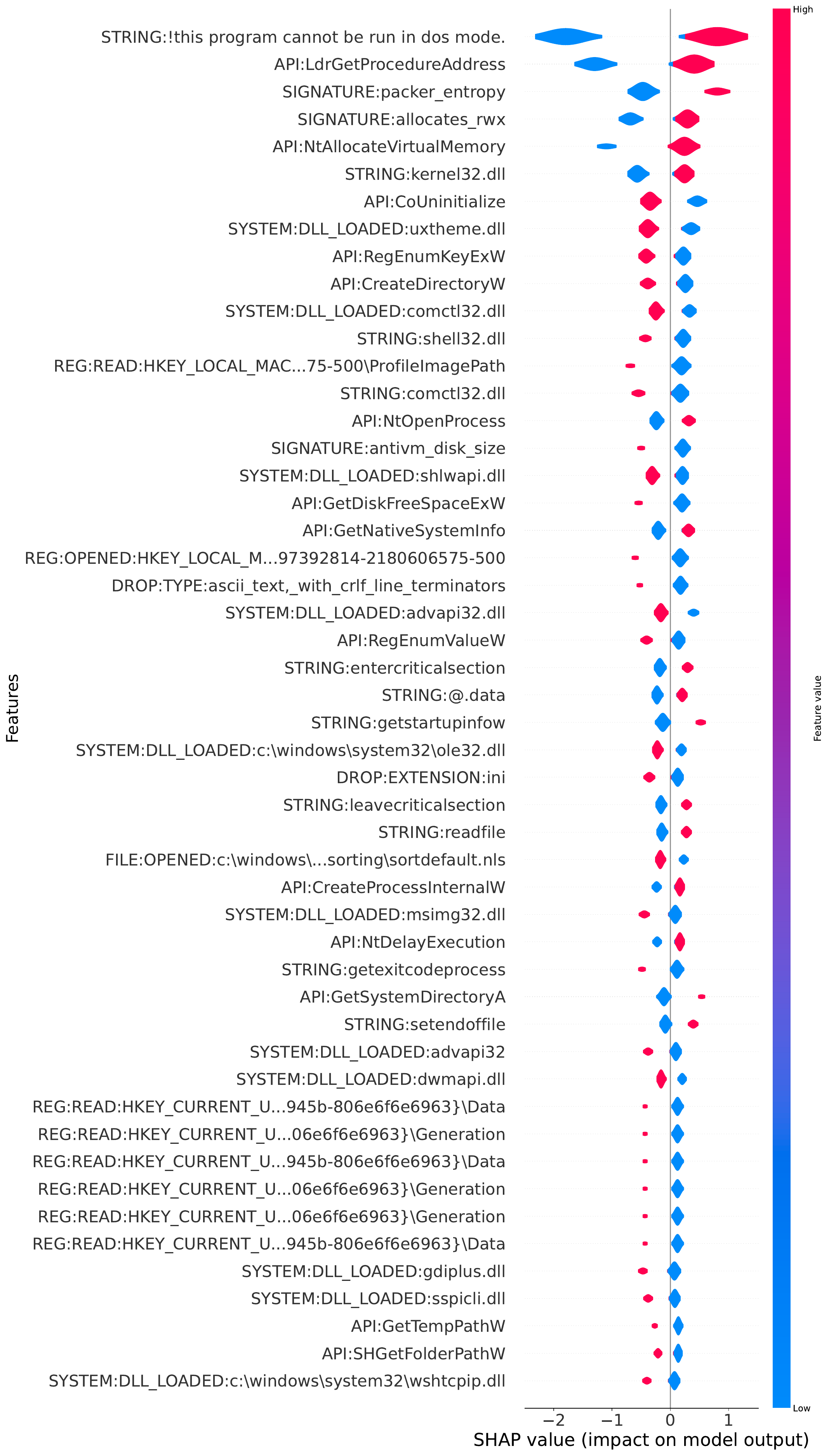}%
\caption{ SHAP violin plot showing the top 50 features ranked by their impact on model output. Wider sections indicate features with greater and more variable influence, while positive and negative values reflect their contribution to the positive and negative class predictions, respectively. }
\label{fig:shap_voilin}
\end{figure*}

Figures \ref{fig:shap_bar} and \ref{fig:shap_voilin} show the SHAP bar plot and SHAP violin plot of the top 50 features that contributed the most to the model's prediction, respectively. It ranks features based on their mean absolute SHAP values, reflecting their impact on the model’s predictions. Closely examining the top 10 features identified by the SHAP analysis provides key insights into how the model distinguishes ransomware from goodware. The highest-ranked feature, \texttt{ (STRING:!this program cannot be run in dos mode)} with a SHAP value of $+1.27$, indicates that the model flags executables designed for modern systems, a common trait of ransomware. Similarly, \texttt{ (API:LdrGetProcedureAddress)} follows by identifying a function call crucial to dynamic linking in Windows. Malware often relies on this technique to load external, hidden code into running processes. This feature signals the model’s sensitivity to dynamic linking, a tactic frequently exploited by ransomware to inject malicious payloads. Together with the first feature, it suggests a strong reliance on identifying signs of obfuscation and covert execution. The model also prioritises \texttt{entropy-based features} such as \texttt{ (SIGNATURE:packer\_entropy)}, which detects packed files commonly used by ransomware to obfuscate their code. 

Additionally, \texttt{ (SIGNATURE:allocates\_rwx)} and \texttt{ (API:NtAllocateVirtualMemory)} reflect the model's ability to identify suspicious memory allocation practices typical of ransomware that manipulates system processes for stealthy execution. The feature \texttt{ (SIGNATURE:allocates\_rwx)} is memory allocation with Read-Write-Execute (RWX) permissions. This tactic is a hallmark of ransomware, as it allows the injection of executable code into running processes. The prominence of this feature signals the model’s sharp focus on detecting suspicious memory behaviours, particularly those indicative of attempts to execute hidden or harmful code. The model continues to build this case with \texttt{ (API:NtAllocateVirtualMemory)}, focusing on memory allocation within a process’s address space. This API call is often associated with malware activities such as runtime payload injection and manipulation of system memory. By highlighting this feature, the model further emphasises its sensitivity to runtime behaviours commonly seen in ransomware.

As we move forward, \texttt{ (STRING:kernel32.dll)} introduces an important system-level component. \texttt{Kernel32.dll} is essential for core Windows functions, and its interaction with software could signal attempts by malware to manipulate system-level operations. This feature enhances the model’s ability to detect interactions with critical system resources, which are frequently targeted by malicious software seeking to gain elevated privileges or maintain persistence.

Furthermore, \texttt{ (API:CoUninitialize)} shifts the focus to the use of COM (Component Object Model) components. Although not inherently malicious, COM is often exploited by malware to execute hidden functions. The model uses this feature to spot unusual interactions with COM objects, an indication that the software might be attempting to mask its true purpose, a strategy commonly adopted by ransomware.

Features like \texttt{ (SYSTEM:DLL\_LOADED:uxtheme.dll)} and \texttt{ (API:CreateDirectoryW)} further indicate the model’s detection of legitimate system behaviours hijacked by ransomware, such as loading system DLLs or creating directories for malicious payloads. 
These top features show that the model successfully identifies malicious activities related to \textbf{dynamic code execution}, \textbf{memory manipulation}, and \textbf{system modifications}, which are characteristic of modern ransomware attacks.


\begin{table}[htpb]
\centering
\caption{Feature group distribution of the top 50 features based on SHAP. The feature groups are: application programming interface (API), registry keys (REG), file operations (FILE), directory operations (DIR), strings (STR), network activities (NET), system operations (SYS), drop operations (DROP), and signatures (SIG). N represents the number of features belonging to that group and \% represents the percentage of each feature group relative to the total features selected by SHAP in the top 50.}
\label{tab:top50_fxgrp}
\begin{tabular}{lcc}
\hline
\textbf{\begin{tabular}[c]{@{}l@{}}Feature \\ Groups\end{tabular}} & \textbf{N}             & \textbf{\%}          \\ \hline
\textbf{API}  & 13 & 26.0 \\
\textbf{REG}  & 8  & 16.0 \\
\textbf{FILE} & 1  & 2.0  \\
\textbf{DIR}  & 0  & 0.0  \\
\textbf{STR}  & 11 & 22.0 \\
\textbf{NET}  & 0  & 0.0  \\
\textbf{SYS}  & 11 & 22.0 \\
\textbf{DROP} & 2  & 4.0  \\
\textbf{SIG}  & 4  & 8.0  \\ \hline
Total                                                              & \multicolumn{1}{l}{50} & \multicolumn{1}{l}{} \\ \hline
\end{tabular}
\end{table}

Furthermore, we analyse the top 50 features based on SHAP into their respective feature groups as shown in Table \ref{tab:top50_fxgrp}. \textbf{API} is the highest-ranking feature group, with 13 features (26\%), indicating the model’s strong reliance on API calls to detect ransomware and goodware dynamic behaviours. This group is critical in identifying how software interacts with the system at runtime, providing insights into potential malicious behaviour. Following closely are the \textbf{STR} and \textbf{SYS} groups, each contributing 11 features (22\%). \textbf{STR} features focus on string patterns, such as file paths or executable names, which can highlight signatures of malicious software. On the other hand, the \textbf{SYS} group detects system-level operations like memory manipulation and interaction with critical system components, which are often exploited by ransomware to gain control or elevate privileges.

The next most frequent group is \textbf{REG} with 8 features (16\%), highlighting the importance of registry interactions in detecting malware persistence and system alterations. \textbf{SIG} follows with 4 features (8\%), aiding in detecting known malicious software through signature-based methods. The \textbf{DROP} group contributes 2 features (4\%), reflecting sensitivity to file drop operations, commonly used by ransomware.

The \textbf{FILE}, \textbf{DIR}, and \textbf{NET} groups contribute the least, with \textbf{FILE} representing 1 feature (2\%) and \textbf{DIR} and \textbf{NET} having no features (0\%), suggesting they play a minor role in the top 50 features.

\begin{tcolorbox}[
  colback=lightash, 
  colframe=black!50!black, 
  title=Result 3: Most important features,
  enhanced jigsaw,
  breakable
]
SHAP analysis shows that API calls, string patterns, system behaviours, and registry keys are the most predictive feature groups, making up over 70\% of the top 50 features. Key features include dynamic linking, memory use, and interactions with critical components. File, directory, and network feature groups contribute little, reflecting the model’s focus on runtime and obfuscation behaviour.
\end{tcolorbox}


\subsection{Model decision analysis (RQ4)}

\subsubsection{Misclassified model decisions}

\begin{figure}[htpb]
\centering 
  \includegraphics[width=0.99\columnwidth]{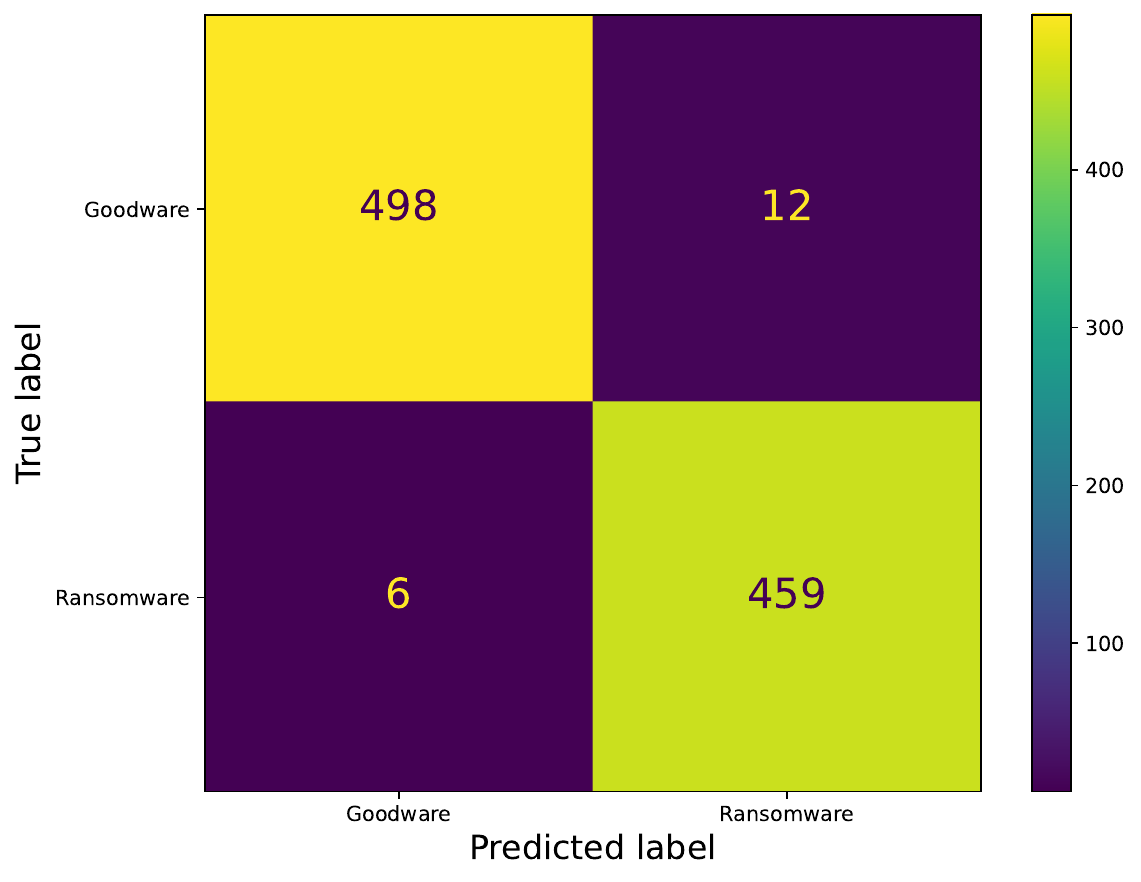}%
\caption{Confusion matrix for the ransomware detection logistic regression model based on the two-stage feature selection technique. The results show that the model correctly classified 498 Goodware and 459 Ransomware instances, with 12 false positives and 6 false negatives. }
\label{fig:confusion_matrix}
\end{figure}

Figure \ref{fig:confusion_matrix} visualises the performance of the Logistic Regression model in classifying goodware and ransomware. The rows represent the true labels, while the columns represent the predicted labels. The matrix indicates that the model correctly identified 498 instances of goodware and 459 instances of ransomware. There were 12 goodware instances misclassified as ransomware (false positives) and 6 ransomware instances misclassified as goodware (false negatives), achieving an accuracy of 98.15\% as shown in Table \ref{tab:rfe_fs}.

Out of the 6 ransomware instances misclassified as goodware, the misclassifications were distributed across three ransomware families: \textbf{Shodi} (4 instances), \textbf{Clop} (1 instance), and \textbf{Delshad} (1 instance). These misclassifications reflect the model's difficulty in distinguishing between these ransomware families. The misclassified instances also span two ransomware types, including \textbf{Crypto} (5 instances) and \textbf{Modern} (1 instance). Also, the misclassified ransomware instances span across different \textbf{first submission years}, with 4 instances first seen in 2021, 1 instance in 2023, and 1 instance in 2024. 

The 12 \textbf{goodware} samples misclassified as \textbf{ransomware} are associated with several \textbf{software sources}, with the majority (10 instances) originating from \textit{Software Informer’s Most Popular} category, and the remaining instances from \textit{Education} (1), and \textit{Business} (1). These misclassified samples include legitimate software programs such as \texttt{pdfcreator}, \texttt{zoominstallerfull}, \texttt{teamspeak3-client}, and \texttt{avast\_secure\_browser\_setup}, among others.

The misclassified samples are also distributed across different \textbf{first submission years}: 2024 (7 instances), 2023 (4 instances), and 2022 (1 instances). This suggests that the model is generally more effective at classifying goodware from earlier years (2022) but struggles more with programs submitted in the last two years (2023 and 2024). The model might encounter challenges due to new software or updates that exhibit behaviour similar to ransomware, leading to false positives, especially for popular software versions.

\begin{figure}[htpb]
\centering 
  \includegraphics[width=0.99\columnwidth]{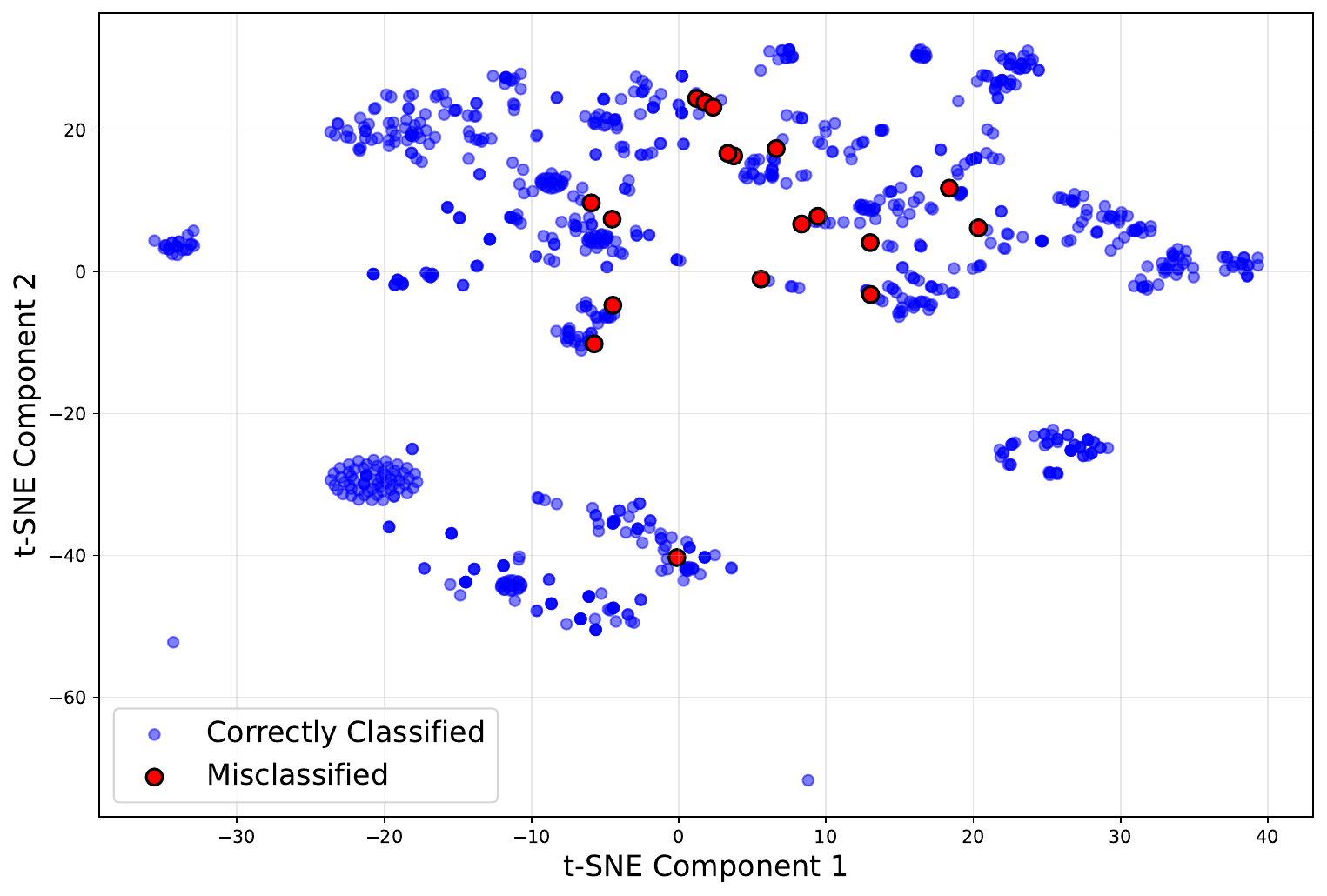}%
\caption{t-SNE projection in 2D with misclassified samples highlighted in red with black borders and correctly classified in blue.}
\label{fig:tsne}
\end{figure}

\begin{figure}[htpb]
\centering 
  \includegraphics[width=0.99\columnwidth]{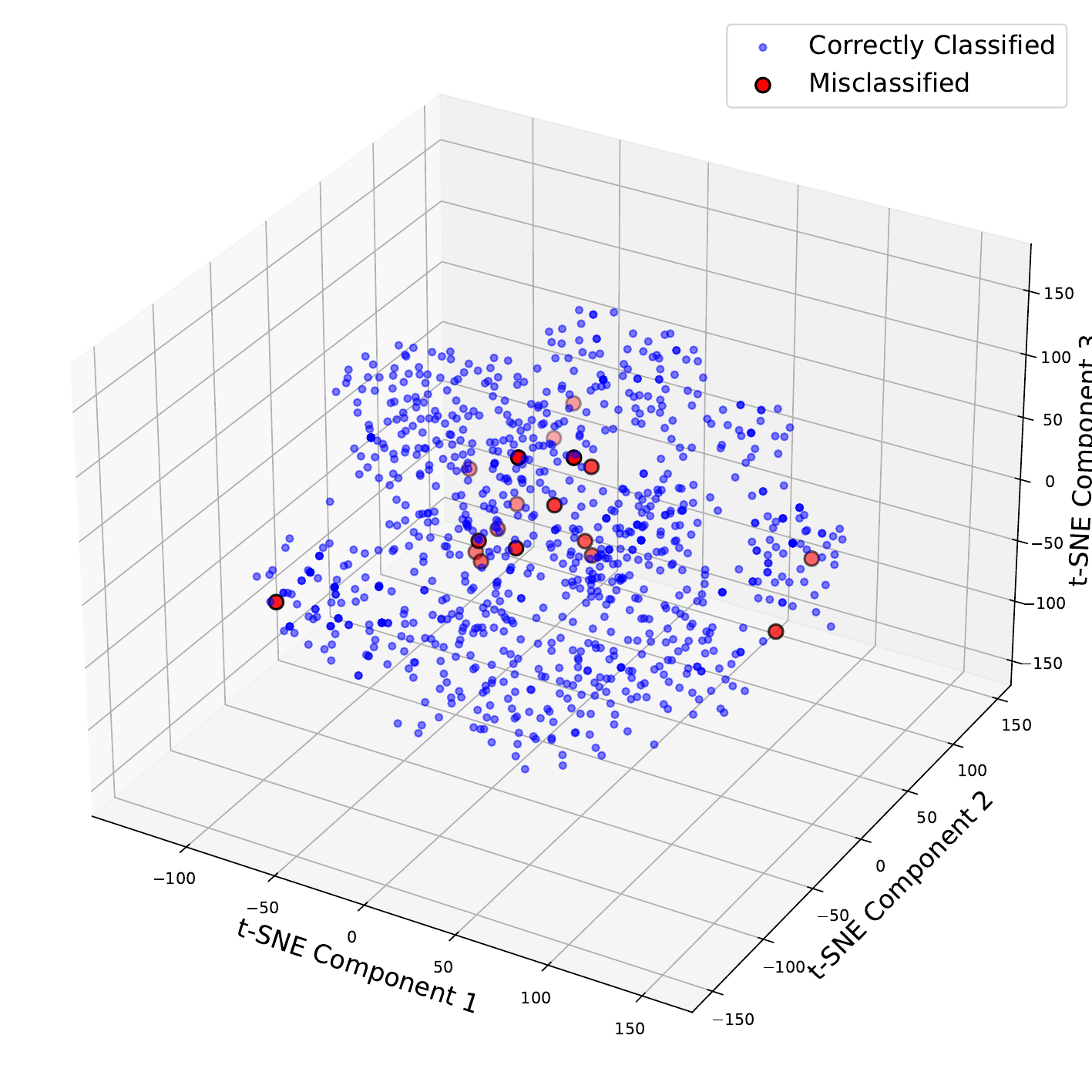}%
\caption{t-SNE projection in 3D with misclassified samples highlighted in red with black borders and correctly classified in blue. }
\label{fig:3dtsne}
\end{figure}

To investigate the decision boundary and understand where the misclassified samples fall in the feature space, we used the dimensionality reduction technique,  t-Distributed Stochastic Neighbor Embedding (t-SNE) to project the high-dimensional data into a 2D and 3D space for visualisation as shown in Figures \ref{fig:tsne} and \ref{fig:3dtsne} respectively. Misclassified samples are highlighted in red with black borders, and correctly classified samples are in blue. As observed in both plots, the correctly classified samples tend to form distinct clusters, indicating that the model can effectively separate different classes in the feature space.
The misclassified samples are scattered near the correctly classified ones, suggesting that the model struggles to differentiate between certain samples. These samples might be ambiguous or fall near the decision boundary of the model.
\begin{tcolorbox}[colback=lightash, colframe=black!50!black, title=Result 4: Misclassified model decision analysis] 
Misclassification analysis reveals that most false positives involve recent, popular goodware exhibiting ransomware-like behaviour, while false negatives cluster around a few ransomware families. t-SNE visualisations show misclassified samples near decision boundaries, indicating ambiguity. These insights expose model weaknesses and inform strategies to improve robustness and interpretability.
\end{tcolorbox}

\subsubsection{Local misclassified instance explanation}

To understand why the model misclassified certain instances, we conducted a local instance analysis using Local Interpretable Model-agnostic Explanations (LIME). This method helps to clarify the reasons behind the model's specific predictions.

\begin{figure*}[]
    \centering
    \includegraphics[width=1\textwidth]{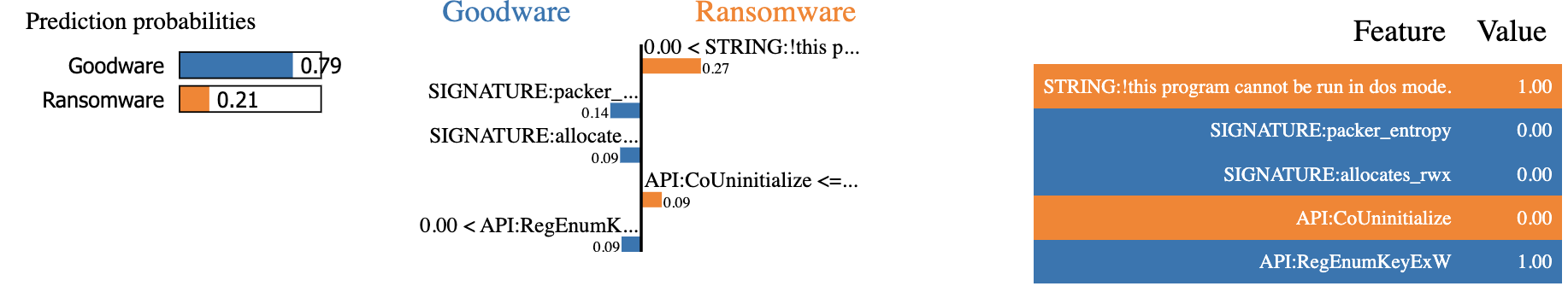}
    \caption{LIME explanation of the model's prediction for a ransomware sample misclassified as goodware. The figure displays the top 5 features and the predicted probabilities for Goodware (79\%) and Ransomware (21\%), along with the contributions of individual features to the model’s decision. Features influencing the Goodware prediction are shown on the left, while those contributing to the Ransomware classification are shown on the right. The feature values for the misclassified sample are also presented.}%
    \label{fig:rw_mis_as_gw1}
\end{figure*}

Figure \ref{fig:rw_mis_as_gw1} shows the LIME explanation for why a ransomware sample was misclassified as goodware. This misclassification is primarily due to the model's reliance on features typically associated with goodware. Specifically, \texttt{Feature (SIGNATURE:packer\_entropy)} strongly favoured goodware, contributing significantly to the probability of classifying the sample as goodware. While \texttt{Feature (STRING:!this program cannot be run in dos mode)} and \sloppy \texttt{Feature (REG:READ:HKEY\_LOCAL\_MACHINE\textbackslash{}SOFTWARE\textbackslash{}Microsoft\textbackslash{}Windows NT\textbackslash{}CurrentVersion\textbackslash{}ProfileList\textbackslash{}S-1-5-21-4114181432- 2397392814-2180606575-500\textbackslash{}ProfileImagePath)} indicated ransomware. The model’s emphasis on goodware-associated features outweighed the signals from ransomware-associated features. This highlights the challenge of distinguishing between the two classes when feature values overlap or when certain features disproportionately influence the model’s decision-making process.

\begin{figure*}[]
    \centering
    \includegraphics[width=1\textwidth]{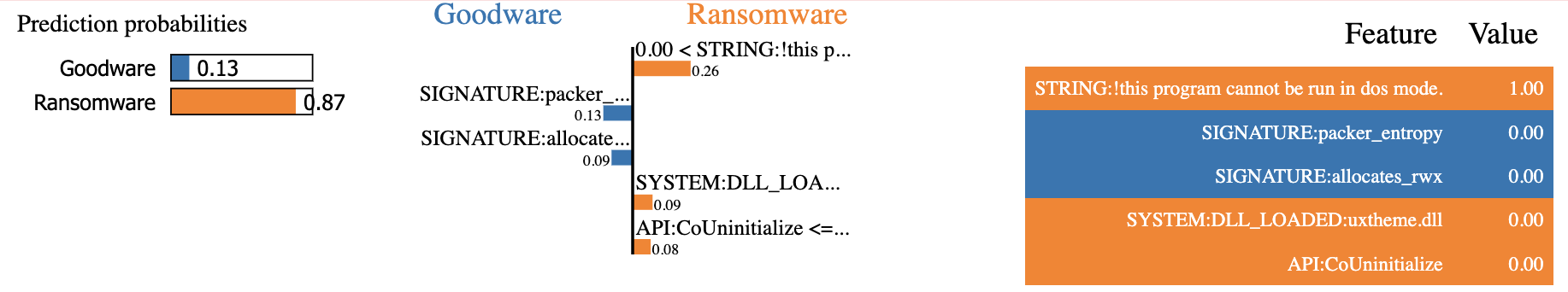}
    \caption{LIME explanation of the model's prediction for a goodware sample misclassified as ransomware. The figure displays the top 5 features and the predicted probabilities for Goodware (13\%) and Ransomware (87\%), along with the contributions of individual features to the model’s decision. Features influencing the Goodware prediction are shown on the left, while those contributing to the Ransomware classification are shown on the right. The feature values for the misclassified sample are also presented.}%
    \label{fig:gw_mis_as_rw1}
\end{figure*}

Similarly, Figure \ref{fig:gw_mis_as_rw1} shows the LIME explanation of a goodware sample misclassified as ransomware. It shows that the model was heavily influenced by \texttt{Feature (STRING:!this program cannot be run in dos mode)}, with a probability of \texttt{0.26}, which is associated with Ransomware. Despite other features like \texttt{Feature (SIGNATURE:packer\_entropy)} and \texttt{Feature (SIGNATURE:allocates\_rwx)} showing benign behaviour, the model's reliance on the strong ransomware-associated features led to the misclassification. 

\begin{tcolorbox}[colback=lightash, colframe=black!50!black, title=Result 5: Local misclassified instance explanation] 
LIME analysis reveals misclassifications often occur when the model overweights a few dominant features. In ransomware samples misclassified as goodware, benign indicators like low entropy suppressed the influence of malicious signals. Conversely, goodware samples were misclassified as ransomware due to strong associations with features commonly seen in malware, such as specific string patterns and registry accesses.
\end{tcolorbox}

\section{Discussion} \label{sec:discussion}
This paper has presented a novel approach to ransomware detection using machine learning. To contextualise our findings and provide a broader understanding of the field, this section critically reviews existing literature on ransomware detection using machine learning, focusing on the efficacy of traditional classifiers, the role of feature selection, the predictive power of different feature groups, and the insights gained from analysing misclassified instances. Also, we discuss the threat to validity and limitations of this study. 

\subsection{Machine learning models for ransomware detection}



\subsubsection{Ransomware-goodware classification}

Our results support prior findings that traditional machine learning classifiers, particularly logistic regression, are highly effective for ransomware-goodware classification when trained on behavioural features. Similar to the EldeRan dataset \citep{sgandurra2016automated}, our logistic regression model achieved strong performance, but with the added benefit of enhanced efficiency through our proposed two-stage feature selection technique. This approach significantly reduced the feature space, from over 6 million to just 483, while maintaining high accuracy, offering a practical and interpretable solution for real-world deployment.

Contrary to earlier studies that favour ensemble methods for ransomware detection \citep{jaya2022dynamic, aljabri2024ransomware}, our evaluation showed that LightGBM performed worst, with lower accuracy and substantially higher computation time. This suggests that simpler models can outperform more complex alternatives when paired with well-chosen features, especially on high-dimensional behavioural data. Moreover, while most previous work focuses solely on predictive performance, our study emphasises the importance of model efficiency and feature reduction, addressing a key limitation in the current literature. Overall, our findings highlight that logistic regression, combined with lightweight feature selection, offers a robust and scalable approach to binary ransomware detection.



\subsubsection{Impact of feature selection}
Effective feature selection is crucial in dynamic ransomware detection due to the high dimensionality of behavioural data. Our two-stage approach (applying MI with a 0.01 threshold followed by RFE) reduced the MLRan dataset from over 6.4 million to just 483 features, significantly improving efficiency and interpretability without sacrificing performance. This builds on prior work using MI \citep{sgandurra2016automated, abbasi2020particle}, but improves upon it by avoiding fixed feature quotas instead refining selection through performance-guided elimination.

While studies such as \citet{moreira2022understanding} and \citet{onwuegbuche2023enhancing} have shown the benefits of reducing feature sets to 300–500 using various techniques, our method achieves similar reductions from a much larger initial space, demonstrating greater scalability. Importantly, our results confirm that the most predictive behaviours lie in API calls, registry, strings, and system operations, echoing findings from previous work. 
Overall, our approach offers a simple yet powerful framework for reducing noise, improving model generalisation, and capturing core ransomware behaviours in large-scale datasets.

\subsubsection{Important features}
Our SHAP analysis highlights API calls, string patterns, system behaviours, and registry activity as the most predictive features for ransomware detection, accounting for over 70\% of the top 50 features. This complements findings by \citet{gulmez2024xran}, who also identified API and registry features as strong ransomware indicators, though they found mutexes and DLLs more relevant in some contexts. Instead of using focused input types, our broader feature set reveals that runtime behaviours, such as dynamic linking and memory allocation, are key discriminators. Overall, our results reinforce that models benefit most from features capturing obfuscation, memory manipulation, and system-level interactions, and demonstrate the value of SHAP for uncovering these behavioural signals.



\subsubsection{Model decision analysis}
Understanding why models misclassify samples is critical for improving detection robustness. Our LIME-based analysis shows that misclassifications often stem from the model’s overreliance on a few dominant features. In false negatives, benign indicators such as low entropy masked ransomware signals, while false positives occurred when benign software exhibited traits commonly associated with malware, like specific strings or registry activity. These findings align with \citet{gulmez2024xran}, who observed that cryptographic APIs and ambiguous DLLs contributed to misclassification, highlighting the nuanced interplay of features rather than their individual presence.
However, our work extends prior efforts by demonstrating how local explanations reveal imbalances in feature influence, even within correctly engineered models. By exposing instances where legitimate behaviours are misinterpreted due to overlapping traits with ransomware, our analysis underscores the need for more discriminative and context-aware features.



\subsection{Threat to validity}

While this study takes deliberate steps to ensure methodological soundness in constructing and evaluating the MLRan dataset, we acknowledge potential threats to validity that, while mitigated, cannot be entirely eliminated.

\subsubsection{Construct validity}  
MLRan is built using behavioural data collected from dynamic analysis in sandboxed environments, which may not perfectly replicate all real-world execution conditions. However, we addressed this by incorporating a diverse set of ransomware and goodware samples and ensuring the sandboxes closely replicate a real-world system. Furthermore, careful labelling and validation steps were taken to reduce misclassification risk, though the evolving nature of malware may still introduce edge cases that challenge labelling frameworks.

\subsubsection{Internal validity}  
Using mutual information followed by recursive feature elimination, the two-stage feature selection approach was designed to enhance model interpretability and reduce overfitting. Although thresholds and model parameters were chosen based on empirical performance and prior literature, we recognise that some subtle features may be filtered out. To address this, we validated performance across multiple models and feature groups, ensuring that the retained features remained representative of key ransomware behaviours.

\subsubsection{External validity}  
While our models and findings are based on the MLRan dataset, which captures a rich and diverse set of ransomware and goodware behaviours, generalisability to unseen environments or malware families may vary. We mitigated this by including recent and varied ransomware types and conducting binary and multiclass evaluations. Nevertheless, further validation on independent datasets would help extend the applicability of the findings.

\subsubsection{Interpretability validity}  
We employed state-of-the-art XAI methods, including SHAP and LIME, to provide insights into model decision-making. Although these techniques offer robust explanations, their assumptions (e.g., feature independence in SHAP, local approximation in LIME) may not fully capture all aspects of model logic. To address this threat to validity, we used global and local explanation methods and cross-referenced feature importance rankings with domain knowledge to ensure interpretative reliability.

In summary, while no empirical study is entirely free from validity concerns, we took rigorous and transparent steps to reduce their impact. The design of MLRan, the careful selection and validation of features, and the application of interpretable machine learning all contribute to the reliability and reproducibility of the presented results.

\section{Conclusion and future work} \label{sec:conclusion}
This study presents a comprehensive framework for advancing behavioural ransomware detection by creating the MLRan ransomware dataset (the largest and most diverse publicly available behavioural ransomware dataset to date) and developing practical open-source tools for advancing ransomware research. We introduced GUIDE-MLRan, a set of structured guidelines to support reproducible, high-quality ransomware dataset construction. We applied these principles to develop MLRan. Covering four major ransomware types and 64 families, and spanning nine behavioural feature groups, MLRan fills a critical gap in existing datasets by offering breadth, depth, and balance.

To facilitate scalable data collection, we enhanced Cuckoo Sandbox through automation scripts for file submission and result sorting, reducing manual effort and improving consistency. Using this infrastructure, we constructed a robust machine learning pipeline that integrates a novel two-stage feature selection method (mutual information followed by recursive feature elimination), which reduced the feature space from over 6.4 million to 483 while maintaining high classification performance. This reduction significantly improves model training time and interpretability. Through global (SHAP) and local (LIME) explainable AI techniques, we further analysed the models’ decision-making, identifying API usage, string patterns, memory operations, and registry behaviours as the most predictive signals of ransomware activity. These findings validate the effectiveness of MLRan and contribute to a deeper understanding of ransomware behaviour.
While our results demonstrate strong model performance in binary classification, they also reveal challenges in multiclass settings and instances of feature ambiguity leading to misclassification. This underscores the need for continued refinement of feature engineering and interpretability tools.

In future work, we will continue our research following several directions. First, we will evaluate model robustness by incorporating adversarial samples in the dataset to study model robustness against evasion techniques. Second, we will develop context-aware models that consider temporal sequences and inter-feature dependencies using graph learning or attention-based architectures. Finally, we encourage the community to use, extend, and benchmark on MLRan, as all datasets, tools, and code are openly released to foster reproducibility and collaboration in this critical domain.



\printcredits

\section*{Declaration of competing interest}
The authors declare that they have no known competing financial interests or personal relationships that could have appeared to
influence the work reported in this paper.

\section*{Acknowledgements}
This work was funded by Research Ireland through the Research Ireland Centre for Research Training in Machine Learning (18/CRT/6183). 

This article was supported in part by Research Ireland Grant 13/RC/2094\_2. 

We also acknowledge Răzvan Maioru for his contribution in analysing a few of the samples using the Cuckoo sandbox.

\section*{Data availability}

The MLRan dataset, along with all associated code and tools, is publicly available at: \url{https://github.com/faithfulco/mlran}.


\clearpage

\appendix

\section{Appendix}

\begin{table*}[h]
\centering
\caption{Model hyperparameters and justifications for each algorithm used in the study, including their value ranges and rationale based on empirical or theoretical considerations.}
\label{tab:hyperparameters_justified}
\resizebox{\textwidth}{!}{%
\begin{tabular}{llll}
\hline
\textbf{Model} &
  \textbf{Hyperparameter} &
  \textbf{Values} &
  \textbf{Justification} \\ \hline
\multirow{3}{*}{\textbf{LightGBM}} &
  num\_leaves &
  {[}15, 31, 63{]} &
  \begin{tabular}[c]{@{}l@{}}Controls model complexity; powers of 2 allow a trade-off between   \\ expressiveness and overfitting.\end{tabular} \\
 &
  n\_estimators &
  {[}50, 100, 200{]} &
  More trees improve stability and performance at the cost of training time. \\
 &
  learning\_rate &
  {[}0.01, 0.1{]} &
  Smaller values converge more smoothly and generalise better. \\ \hline
\multirow{2}{*}{\textbf{\begin{tabular}[c]{@{}l@{}}Decision \\ Tree\end{tabular}}} &
  max\_depth &
  {[}None, 10, 20{]} &
  Restricts depth to prevent overfitting; None allows full growth. \\
 &
  min\_samples\_split &
  {[}2, 5, 10{]} &
  Higher values reduce overfitting by requiring more samples per split. \\ \hline
\multirow{3}{*}{\textbf{\begin{tabular}[c]{@{}l@{}}Logistic \\ Regression\end{tabular}}} &
  C &
  {[}0.01, 0.1, 1, 10{]} &
  \begin{tabular}[c]{@{}l@{}}Inverse regularisation strength; wider range allows control over model \\ complexity.\end{tabular} \\
 &
  penalty &
  {[}'l2'{]} &
  Standard regularisation for numerical stability and generalisation. \\
 &
  solver &
  \begin{tabular}[c]{@{}l@{}}{[}'lbfgs' (multiclass), \\ 'liblinear' (binary){]}\end{tabular} &
  Solvers selected for efficiency and compatibility with classification type. \\ \hline
\multirow{3}{*}{\textbf{\begin{tabular}[c]{@{}l@{}}Random\\ Forest\end{tabular}}} &
  n\_estimators &
  {[}50, 100, 200{]} &
  Larger ensembles reduce variance; diminishing returns beyond a point. \\
 &
  max\_depth &
  {[}None, 10, 20{]} &
  Controls overfitting; allows shallower trees for better generalisation. \\
 &
  min\_samples\_split &
  {[}2, 5{]} &
  Prevents overly specific branches that may not generalise well. \\ \hline
\multirow{3}{*}{\textbf{\begin{tabular}[c]{@{}l@{}}Extra \\ Trees\end{tabular}}} &
  n\_estimators &
  {[}50, 100, 200{]} &
  Similar to Random Forest; more trees enhance robustness. \\
 &
  max\_depth &
  {[}None, 10, 20{]} &
  Regulates tree complexity and training time. \\
 &
  min\_samples\_split &
  {[}2, 5{]} &
  Enforces minimum data per split; helps mitigate overfitting. \\ \hline
\end{tabular}%
}
\end{table*}

\bibliographystyle{cas-model2-names}

\bibliography{cas-dc-template}


\bio{figs/authors/faithfulco}
Faithful Chiagoziem Onwuegbuche is a PhD candidate in Machine Learning at the SFI Center for Research Training in Machine Learning (ML-Labs) at University College Dublin (UCD), Ireland. He is also affiliated with the SFI Research Centre for Software (LERO). His research interests lie at the intersection of artificial intelligence and blockchain technology, with a focus on their applications in finance and cybersecurity. His ultimate goal is to develop AI systems capable of securing critical infrastructure, detecting financial fraud, and promoting financial inclusion. Prior to his PhD, he earned two master's degrees: a Master of Science in Financial Technology (FinTech) with Distinction from the University of Stirling, UK, and a Master of Science in Financial Mathematics from the Pan African University, Kenya, as a Commonwealth Shared Scholar and African Union Scholar, respectively. He has professional experience as a data scientist, machine learning researcher, and lecturer. You can read more about him at: \url{https://faithfulco.github.io/}.
\endbio

\bio{figs/authors/adelodun}
Sunday Adelodun holds a B.Sc. in Electrical and Electronic Engineering from the University of Ibadan, Nigeria. He is an Embedded Systems Engineer specializing in firmware development, embedded computing, and system architecture. His research interests include embedded systems design, IoT, and Edge AI, with a focus on optimizing performance and efficiency in resource-constrained environments, including systems leveraging embedded operating systems.
\endbio

\bio{figs/authors/Anca}
Dr. Anca Jurcut is an Assistant Professor in the School of Computer Science, University College
Dublin (UCD), Ireland, since 2015. She received a BSc in Computer Science and Mathematics from
West University of Timisoara, Romania in2007 and a PhD in Security Engineering from the
University of Limerick (UL), Ireland in 2013 funded by the Irish Research Council for Science
Engineering and Technology. She worked as a postdoctoral researcher at UL as a member of the
Data Communication Security Laboratory and as a Software Engineer in IBM in Dublin, Ireland
in the area of data security and formal verification. Dr. Jurcut research interests include
Security Protocols Design and Analysis, Automated Techniques for Formal Verification, Network
Security, Attack Detection and Prevention Techniques, Security for the Internet of Things, and
Applications of Blockchain for Security and Privacy. Dr. Jurcut has several key contributions in
research focusing on detection and prevention techniques of attacks over networks, the
design and analysis of security protocols, automated techniques for formal verification, and
security for mobile edge computing (MEC). More Info: \url{https://people.ucd.ie/anca.jurcut}.
\endbio

\bio{figs/authors/liliana}
Liliana Pasquale received her PhD from Politecnico di Milano (Italy) in 2011. She is an Associate Professor at University College Dublin (Ireland) and a funded investigator at Lero - the SFI Research Centre for Software. Her research interests include requirements engineering and adaptive systems, focusing on security, privacy, and digital forensics. She has served in the Program and Organizing Committee of prestigious software engineering conferences, such as ICSE, FSE, ASE, RE. She is an associate editor of the IEEE TSE journal, department editor of the IEEE Security \& Privacy Magazine and a member of the review board of the ACM  TOSEM journal.
\endbio

\end{document}